\documentclass{elsart}

\usepackage{amsmath,amssymb,graphicx,bm}

\newcommand{\beqn}{\begin{equation}}
\newcommand{\eeqn}{\end{equation}}
\newcommand{\bea}{\begin{eqnarray}}
\newcommand{\eea}{\end{eqnarray}}
\newcommand{\ba}{\begin{align}}
\newcommand{\ea}{\end{align}}
\newcommand{\m}{|}
\newcommand{\lm}{\Lambda}

\newcommand{\vlowk}{V_{{\rm low}\,k}}
\newcommand{\vlowkherm}{\overline{V}_{{\rm low}\,k}}
\newcommand{\tlowk}{T_{{\rm low}\,k}}
\newcommand{\tlowkherm}{\overline{T}_{{\rm low}\,k}}
\newcommand{\veff}{V_{\rm eff}}
\newcommand{\teff}{T_{\rm eff}}
\newcommand{\vnn}{V_{\rm NN}}
\newcommand{\tnn}{T_{\rm NN}}
\newcommand{\glam}{G^{\Lambda}_0}
\newcommand{\glambar}{\overline{G^{\Lambda}_0}}

\newcommand{\fmi}{\, \text{fm}^{-1}}

\newcommand{\la}{\langle}
\newcommand{\ra}{\rangle}

\newcommand{\wt}{\widetilde}

\newcommand{\chip}{|\chi_{p}\ra}
\newcommand{\chipt}{\la\widetilde \chi_{p}|}
\newcommand{\ddlamfsq}{\frac{d}{d\Lambda}[f^2(p)]}
\newcommand{\ddlamfsqdisc}{\frac{d}{d\Lambda}[f^2_p]}
\newcommand{\openone}{\leavevmode\hbox{\small1\normalsize\kern-.33em1}}

\begin{document}

\begin{frontmatter}

\title{Low-momentum interactions \\
with smooth cutoffs}

\author{S.K.\ Bogner}$^1$,
\ead{\\bogner@mps.ohio-state.edu}
\author{R.J.\ Furnstahl}$^1$,
\ead{\\furnstahl.1@osu.edu}
\author{S.\ Ramanan}$^1$ and
\ead{\\suna@mps.ohio-state.edu}
\author{A.\ Schwenk}$^{2,3}$
\ead{\\schwenk@triumf.ca}
\address{$^1$Department of Physics,
The Ohio State University, Columbus, OH\ 43210\\
$^2$Department of Physics,
University of Washington, Seattle, WA\ 98195-1560\\
$^3$TRIUMF, 4004 Wesbrook Mall, Vancouver, BC, Canada, V6T 2A3}


\begin{abstract}
Nucleon-nucleon potentials evolved to low momentum, which
show great promise in few- and many-body calculations, have
generally been formulated with a sharp cutoff on relative
momenta. However, a sharp cutoff has technical disadvantages
and can cause convergence problems at the 10--100\,keV level
in the deuteron and triton.
This motivates using smooth momentum-space
regulators as an alternative.
We generate low-momentum interactions with smooth cutoffs
both through energy-independent renormalization group methods
and using a multi-step process based on the Bloch-Horowitz
approach. 
We find greatly improved convergence for calculations of 
the deuteron and triton binding energies in a harmonic oscillator basis 
compared to results with a sharp cutoff.
Even a slight evolution of chiral effective field theory interactions to
lower momenta is beneficial.
The renormalization group preserves the long-range
part of the interaction, and consequently the
renormalization of long-range operators, such as the quadrupole
moment, the radius and $\la 1/r \ra$, is small. This demonstrates
that low-energy observables in the deuteron are reproduced without
short-range correlations in the wave function.
\end{abstract}

\end{frontmatter}
\maketitle

\section{Introduction}
\label{sect:intro}

Internucleon potentials with variable momentum cutoffs, known
generically as ``$\vlowk$,'' show great promise for few- and 
many-body 
calculations~\cite{Vlowk1,Vlowk2,VlowkRG,Vlowk3N,Bogner_nucmatt,VlowkSM1,%
VlowkSM2}.
Changing the cutoff leaves observables unchanged by construction, but
shifts contributions between the potential and the
sums over intermediate states in loop integrals.
These shifts can weaken or largely eliminate sources of non-perturbative
behavior such as strong short-range repulsion or the tensor 
force~\cite{Bogner:2006tw}. 
An additional bonus is that the corresponding three-nucleon interactions
become perturbative at lower cutoffs~\cite{Vlowk3N}.
As a result, it is found in practice 
that few- and many-body calculations 
can be greatly simplified or converge more rapidly by lowering 
the cutoff.
This has been observed with few-body variational 
methods~\cite{Bogner:2005fn,Bogner:2006a}, the coupled-cluster 
approach~\cite{VlowkCC},
and for nuclear matter~\cite{Bogner_nucmatt}. 

Low-momentum nucleon-nucleon interactions
were originally derived and constructed in energy-independent form
using model-space methods
(such as Lee-Suzuki~\cite{Vlowk1,Vlowk2} or Okubo~\cite{Epelbaum})
applied in momentum space.
These approaches define orthogonal subspaces with projection operators
$P$ and $Q$, such that $P + Q = 1$ and $P Q = Q P = 0$.
In momentum space, the latter condition implies
a sharp cutoff $\Lambda$
in (relative) momentum, so that $P$-space integrals run
from 0 to $\Lambda$, 
while $Q$-space integrals run from $\Lambda$ to $\infty$
(or to a large ``bare'' cutoff). 
Subsequently, this $\vlowk$ construction was shown to be equivalent to a
Renormalization Group (RG) treatment, derived by requiring cutoff independence 
of the half- or fully-on-shell $T$ matrix~\cite{VlowkRG}.
With a sharp cutoff, the equations take a particularly simple form and
two-body observables are preserved for all momenta up to the cutoff. 

\begin{figure}[t]
\begin{center}
\includegraphics*[width=3.2in]{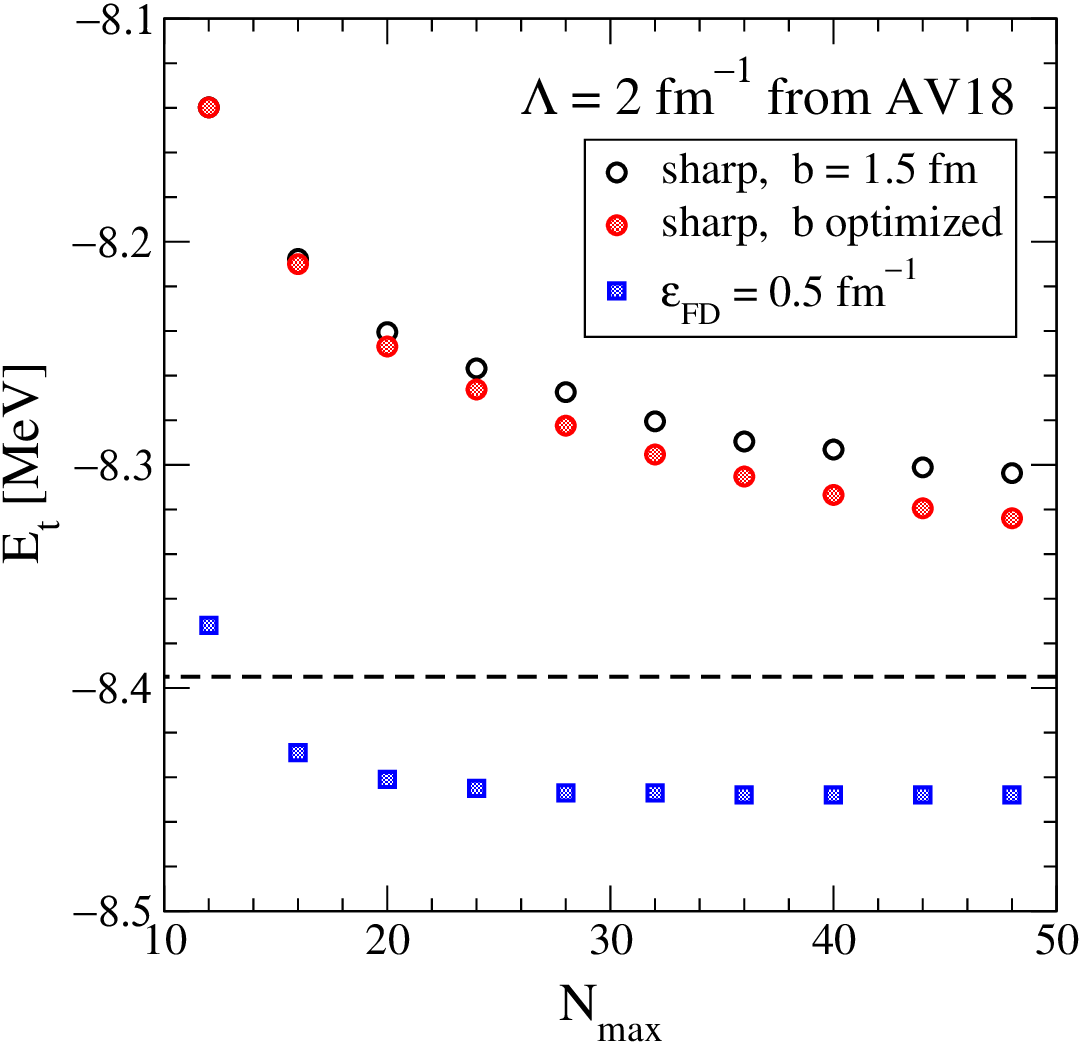}
\end{center}
\caption{The triton binding energy $E_t$ calculated from a direct
diagonalization in a harmonic oscillator basis of the low-momentum
Hamiltonian derived from the
Argonne $v_{18}$ potential~\cite{AV18} with cutoff $\Lambda = 
2\,\mbox{fm}^{-1}$, as a function of the size of the oscillator space
($N_{\rm max} \, \hbar \omega$ excitations).
The open circles are calculated
with a sharp cutoff for a fixed oscillator parameter $b$ while the
filled ones correspond to optimizing $b$ at each $N_{\rm max}$.
The dashed line indicates the exact Faddeev result 
using the sharp-cutoff interaction~\cite{Vlowk3N},
and shows the slow convergence of the diagonalization at the
100\,kev level.
The squares are for a smooth Fermi-Dirac regulator, Eq.~(\ref{eq:fdreg}),
that solves the convergence problem.}
\label{fig:triton}
\end{figure}

However, a sharp cutoff also leads to cusp-like behavior for the 
interaction (close to the cutoff)
in some channels and for the deuteron wave function, which becomes increasingly
evident as the cutoff is lowered below $2 \fmi$. In some applications,
this leads to slow convergence, for example at the 
10--100\,keV level in few-body calculations using harmonic oscillator bases
(see Fig.~\ref{fig:triton}).
The reduction of the repulsive short-range interaction 
simultaneously reduces short-range
correlations in the wave functions, which means that variational calculations
can be effective with much simpler trial ans\"atze~\cite{Bogner:2005fn}.
One would expect that such calculations for the deuteron and the triton,
which are particularly low-energy bound states, should show improvement
for cutoffs well below $2 \fmi$, but instead a degradation
was observed in Ref.~\cite{Bogner:2005fn}.
This result was attributed to the use of sharp cutoffs and
a preliminary study~\cite{Bogner:2006a} showed that these problems
are alleviated by using a smooth-cutoff low-momentum interaction.
In this paper, we verify this conclusion and 
explore in detail the construction and application 
of $\vlowk$ interactions with smooth cutoffs.
 
While smooth cutoffs seem incompatible with methods requiring $PQ = 0$,
it is not a conceptual problem for the RG approach.
Indeed, there is an appreciable literature on smooth-cutoff regulators
for applications of the functional or exact RG~\cite{Liao2000}.
The functional RG
keeps invariant the full generating functional, which
translates into preserving all matrix elements of the inter-nucleon
$T$ matrix.
While this straightforwardly leads to RG equations, it also
implies an energy-dependent interaction, which is undesirable for 
practical few- and many-body calculations.
We resolve this conflict in Sect.~\ref{sect:EdepRG} by 
constructing a low-momentum, energy-independent interaction
with smooth cutoffs in three steps:
\begin{enumerate}
\item Evolve a large-cutoff potential to a lower, smooth cutoff while
preserving the \emph{full} off-shell $T$ matrix. This generates an 
energy-dependent low-momentum interaction.
\item Convert the energy dependence to momentum dependence, 
which results in a non-hermitian smooth-cutoff 
interaction.
\item Perform a similarity transformation to hermitize the low-momentum
interaction.
\end{enumerate}
Along with the possibility of many different functional forms for the
smooth regulators,
various hermitization schemes are possible, which reflects the general
freedom in low-energy effective theories~\cite{Lepage}.

The freedom in defining low-energy
potentials has consequences in practical
calculations.
This is already evident in Fig.~\ref{fig:triton}, where the 
(converged) triton binding
energy for a sharp cutoff is 50\,keV less than for a particular
Fermi-Dirac regulator.  The difference reflects the different contribution
from the (neglected) short-range three-body force.  
Just as changing a cutoff with a sharp regulator moves one along a 
Tjon line~\cite{Vlowk3N}, we expect that
changing the form of the regulator and the hermitization scheme at fixed
cutoff also does. 
Thus, each combination of regulator (specified by one or more ``sharpness''
parameters) and hermitization scheme will have
different corresponding three-body (and higher many-body) interactions.
 
In Sect.~\ref{sect:EindepRG}, we derive alternative
energy-independent RG equations that are generalizations from
sharp cutoffs. This approach has the advantage of a one-step construction,
but accurate solutions of the resulting RG equations for low-momentum
cutoffs are technically more challenging. The low-momentum interactions
with smooth cutoffs, with different regulator types 
and different hermitization 
schemes, are applied in the two-body sector 
and to calculations of the triton in Sect.~\ref{sect:2b}.
We test the convergence properties and document the distortions of
various combinations. 
In most cases we derive the low-momentum interactions starting from chiral
effective field theory (EFT)
potentials at N$^3$LO~\cite{N3LO,N3LOEGM}, but also
use the Argonne $v_{18}$ potential~\cite{AV18} for comparison. 
We summarize our conclusions in Sect.~\ref{sect:concl} and give an outlook 
for future applications.


\section{Smooth cutoff interactions via an energy-dependent RG}
\label{sect:EdepRG}

Our goal is to construct a smooth cutoff version of the
energy-independent and hermitian low-momentum interaction $\vlowk$.
In this section, we describe a three-step method that
utilizes an energy-dependent RG equation to lower the cutoff, followed
by two transformations that remove the energy dependence and
the resulting non-hermiticity in the interaction. 
In the next section, we derive an equivalent method that
uses a hermitian and energy-independent RG equation to construct the 
low-momentum interaction in one step.

First, we derive the RG equation for an energy-dependent low-momentum
interaction $\veff(E)$, which is cut off by smooth regulators.
It is convenient and efficient for numerical calculations to
define the partial-wave
interaction $\veff$ and the corresponding $\teff$ matrix in 
terms of a reduced potential $v$ and a reduced $t$ matrix as
\bea
\veff(k',k;E) &=& f(k') \, v(k',k;E) \, f(k) \,, \\[2mm]
\teff(k',k;E) &=& f(k') \, t(k',k;E) \, f(k) \,,
\eea
where $f(k)$ is a smooth cutoff function satisfying
\beqn
f(k) \stackrel{k \ll \Lambda}{\longrightarrow} 1
\qquad \text{and} \qquad
f(k) \stackrel{k \gg \Lambda}{\longrightarrow} 0 \,.
\label{eq:regulator}
\eeqn
The regulator functions are the same in each partial wave and possible
choices are discussed in Sect.~\ref{sect:2b}.

Given the low-momentum interaction $\veff(E)$, the reduced fully-off-shell 
$t$ matrix satisfies the Lippmann-Schwinger equation,
\beqn
t(k',k;E) = v(k',k;E) + \frac{2}{\pi}\int_{0}^{\infty} p^2dp \,
\frac{v(k',p;E) \, f^2(p) \, t(p,k;E)}{E-p^2} \,,
\eeqn
where we use units with $\hbar = c = m = 1$ and 
a principal value integral is implicit here and 
in the following. Note that the 
smooth cutoff is on the loop momentum but not on external momenta, and 
that $v$ and $f$ are cutoff dependent.
We impose that the fully-off-shell $t(k',k; E)$ is independent of the 
cutoff, $dt(k',k;E)/d\Lambda = 0$. This leads to an energy-dependent 
RG equation for the change in the 
reduced low-momentum interaction $v(k',k;E)$
as the cutoff is lowered. In operator form, we have $t=v\,(1 + \glam\,t)$ and
thus $v=t\,(1 + \glam\,t)^{-1}$. With $d U^{-1} = - U^{-1} \, dU \, U^{-1}$,
we obtain (see also Ref.~\cite{Birse:1998dk})
\beqn
\frac{dv}{d\Lambda} = - t\,(1 + \glam\,t)^{-1} \, \frac{d\glam}{d\Lambda} \,
t\,(1 + \glam\,t)^{-1} = - v \, \frac{d\glam}{d\Lambda} \, v \,.
\eeqn

Restoring the momentum and energy arguments, the energy-dependent RG 
equation reads
\beqn
\frac{d}{d\Lambda} \, v(k',k;E) = \frac{2}{\pi} \int_{0}^{\infty} p^2dp \,
\frac{v(k',p;E) \, \ddlamfsq \, 
 v(p,k;E)}{p^2-E} \,.
\label{EdepRGE}
\eeqn
For the implementation, we discretize the momentum space on a set of 
Gaussian mesh-points, which results in coupled RG equations that can 
be numerically integrated using standard methods. It is convenient to 
define discretized plane-wave states as $|\bar{k}\rangle\equiv k
\sqrt{2w_k/\pi}\,|k\rangle$, 
where $w_k$ are the Gauss-Legendre weights. The new 
basis states are normalized as $\langle\bar{k}|
\bar{p}\rangle = \delta_{k,p}$.
Taking matrix elements of $v(E)$ between the discretized 
plane-wave states gives 
$v_{k',k;E} \equiv \frac{2}{\pi} \, k'k\sqrt{w_{k'}w_k}\, v(k',k;E)$, and 
Eq.~(\ref{EdepRGE}) takes the simple matrix form
\beqn
\frac{d}{d\Lambda} \, v_{k',k;E} = \sum\limits_p \, v_{k',p;E} \:
\frac{\ddlamfsqdisc}{p^2-E} \: v_{p,k;E} \,.
\eeqn
Here and in the following, subscripts denote discrete momentum labels.
In addition, it is convenient to convert the 
principal value integration to a normal integration using
the standard principal-value subtraction method~\cite{NR}.

If we take a nucleon-nucleon (NN) potential model $\vnn$ as the 
large-cutoff initial condition and numerically 
integrate the RG equation, the resulting energy-dependent 
$v(k',k;E)$ preserves 
the fully-off-shell $\tnn$ for \emph{all} external momenta and energies, 
$t(k',k;E)=\tnn(k',k;E)$. Therefore,
$\veff(k',k;E)=f(k') \, v(k',k;E) \, f(k)$ preserves the low momentum 
fully-off-shell $\tnn$ matrix up to factors of the smooth cutoff function, 
$\teff(k',k;E)=f(k') \, \tnn(k',k;E) \, f(k)$.

In the second step, we convert the energy dependence of $\veff(E)$ to momentum 
dependence by using a method similar to field redefinitions. For this
purpose, we introduce an energy-independent (but non-hermitian) $\vlowk(k',k)$ 
that reproduces the half-on-shell $\teff$ matrix (and hence wave functions) 
as $V_{\rm eff}(k',k;E)$,
\beqn
\la k'| \teff(p^2) | p \ra = \la k' | \veff(p^2) | \chi_p \ra
\equiv \la k' | \vlowk | \chi_p \ra \,,
\eeqn
where $\chip$ are the eigenstates of the energy-dependent 
Hamiltonian $H_{\rm eff}(p^2) = T + \veff(p^2)$ with relative
kinetic energy $T$. Using the completeness 
of the interacting eigenstates, this leads to
\beqn
\vlowk(k',k) = \int_{0}^{\infty} p^2dp \: \frac{2}{\pi} 
\int_{0}^{\infty} p'^2 dp' \, \veff(k',p';p^2) \,
\chi_p(p') \, \widetilde{\chi}_p^*(k) \,.
\label{VchiEq}
\eeqn
The energy dependence of $H_{\rm eff}(p^2)$ necessitates the use of 
bi-orthogonal complement vectors $\chipt$ in the 
completeness relation, $\int p^2 dp\, \chip\chipt = \openone$.
If bound states are present, the integral 
over the continuous scattering states includes a summation over bound 
states as well. 

For the numerical solution of Eq.~(\ref{VchiEq}), we first
obtain the eigenstates of the energy-dependent $H_{\rm eff}$ by 
self-consistently diagonalizing the discretized eigenvalue equation
\beqn
\sum_k \bigl( \delta_{k'k} \, k^2 + V^{\rm eff}_{k',k;E_p} \bigr)
\chi_{k,p} = E_p \, \chi_{k',p} \,.
\eeqn   
 
Next, the discretized bi-orthogonal complement
vectors are obtained from the matrix equation
$\sum_p \chi_{k,p} \, \widetilde{\chi}_{k',p}
= \delta_{k,k'}$. Finally, Eq.~(\ref{VchiEq}) 
is evaluated by simple matrix multiplication 
\beqn
V^{{\rm low}\,k}_{k',k} = \sum\limits_{p,p'} \, V^{\rm eff}_{k',p';E_p} 
\, \chi_{p',p} \, \widetilde{\chi}_{k,p} \,.
\eeqn

This procedure is straightforward in
practice. The only subtlety arises when the discretization of the defining
equations for the bi-orthogonal complement vectors is such that
some of the $\chip$ vectors appear to be linearly dependent. In this
case, singular-value-decomposition methods are helpful to solve the singular
system of equations by setting to zero the problematic small singular
values when taking the inverse of the $\chi_{k,p}$ matrix~\cite{NR}.
The resulting energy-independent $\vlowk$ is non-hermitian,
and is identical to the solution of the energy-independent RG equation derived 
from half-on-shell $t$ matrix equivalence in the next section.

The method can be improved further by integrating the
energy-dependent RG equation,
Eq.~(\ref{EdepRGE}), formally instead of a numerical integration.
That is, using  $d v^{-1} = - v^{-1} \, dv \, v^{-1} = d \glam$, 
we obtain $v^{-1}
- \vnn^{-1} = \glam - G_0$, and we recover the Bloch-Horowitz equation 
generalized to a smooth cutoff,
\beqn
v(k',k;E) = \vnn(k',k) + \frac{2}{\pi} \int_{0}^{\infty} p^2dp \,
\frac{\vnn(k',p) \, [1-f^2(p)] \, v(p,k;E)}{E-p^2} \,.
\label{smoothBHE}
\eeqn 
Converting the RG equation to this integral equation and solving
by matrix methods 
speeds up the calculation considerably and reduces the accumulation of 
numerical errors.\footnote{One recovers the folded diagram series, if one 
applies the same trick to the energy-independent RG equation discussed 
in the next section. However, this series does not sum to a 
simple linear integral equation.}

Finally, we note that the initial energy-dependent RG equation is not
needed if one starts directly from the smooth-cutoff generalization of
the Bloch-Horowitz equation. For this purpose, we separate the free
two-nucleon propagator into a smooth-cutoff low-momentum $\glam$ and a
high-momentum part $\glambar$. 
The Lippmann-Schwinger equation then reads 
$\tnn(E)=\vnn+\vnn\,(\glam(E)+\glambar(E))\,\tnn(E)$, and a rearrangement 
gives the fully-off-shell $\tnn(E)$ matrix with only low-momentum 
propagators,
\beqn
\tnn(E) = v(E) + v(E) \, \glam(E) \, \tnn(E) \,, 
\eeqn
where $v(E)$ is the solution to the smooth-cutoff Bloch-Horowitz equation, 
Eq.~(\ref{smoothBHE}).

\subsection{Hermitian low-momentum interactions}

In the third step, we remove the non-hermiticity of $\vlowk$ by a 
similarity transformation that orthogonalizes the set of eigenvectors
$\{ \chip \}$ of the non-hermitian Hamiltonian
$H_{{\rm low}\,k} = T + \vlowk$. Following Holt {\it et al.}~\cite{Holt}, 
we define a transformation $Z$ by
\beqn
Z \chip = \, |\xi_p \ra \qquad \text{and} \qquad
\la \xi_p | \xi_{p'} \ra = \delta_{pp'} \,,
\label{ztrans}
\eeqn
where the Kronecker delta normalization implies the discretization
procedure of the previous section has been carried out.
The hermitian low-momentum interaction is then given by
\beqn
\vlowkherm = Z \, H_{{\rm low}\,k} \, Z^{-1} - T \,.
\eeqn

There is not a unique choice for the transformation $Z$, which is a reflection
of the
general freedom in low-energy effective theories. In Ref.~\cite{Holt}, it 
was shown how several common hermitization methods correspond to different 
choices for $Z$. For example, within the 
Lee-Suzuki framework~\cite{LS1,LS2} for deriving 
effective interactions (which implies a sharp cutoff 
corresponding to orthogonal $PQ=0$ projection operators), 
the eigenstates of the non-hermitian $\vlowk$ obey
\beqn
\la \chi_p | P + \omega^{\dagger} \omega |\chi_{p'} \ra = \delta_{pp'} \,,
\label{nonhermmetric}
\eeqn
where $\omega = Q \omega P$ is the wave operator that 
parameterizes the Lee-Suzuki
decoupling transformation. 
Identifying $Z^{\dagger}Z = P+\omega^{\dagger}\omega$
defines a class of valid transformations for $Z$.  For example, setting 
$Z=\sqrt{P+\omega^{\dagger}\omega}$ corresponds to the hermitization procedure 
of Okubo~\cite{Okubo} and Okamoto and Suzuki~\cite{Okamoto}. 
Alternatively, one can perform a
Cholesky factorization of the symmetric and positive-definite operator 
$LL^{\dagger}=P+\omega^{\dagger}\omega$, where $L$ is the lower-triangular 
Cholesky matrix. Then, $Z=L^{\dagger}$ corresponds to the 
hermitization method of Andreozzi~\cite{Andreozzi}. 
Another method discussed by Holt {\it et al.} is a Gram-Schmidt 
orthogonalization
to construct the set of $\{ |\xi_p \ra \}$ directly from
$\{ \chip \}$. This allows $Z$ to be calculated directly 
from Eq.~(\ref{ztrans}).
Even within the Gram-Schmidt method, $Z$ is not unique, due to the freedom 
in choosing the starting vector. All hermitization methods result 
in low-momentum interactions $\vlowkherm$ that preserve the low-momentum
fully-on-shell $\tnn$ matrix, up to factors of the 
regulator function $\tlowk(k,
k;k^2)=f^2(k)\,\tnn(k,k;k^2)$, and the deuteron binding energy.
The corresponding three-body interactions will differ, however, as
discussed below.

The Gram-Schmidt method can be applied directly to the smooth-cutoff $\vlowk$. 
As in Ref.~\cite{Holt}, the orthogonal 
basis $\{ | \xi_p \ra \}$ is constructed via
\bea
| \xi_{1} \ra &=& Z_{11} | \chi_{1} \ra \nonumber \\[2mm]
| \xi_{2} \ra &=& Z_{21} | \chi_{1} \ra
+ Z_{22} | \chi_{2}\ra \\[2mm]
|\xi_{3} \ra &=& Z_{31} | \chi_{1} \ra
+ Z_{32} | \chi_{2} \ra + Z_{33} | \chi_{3} \ra \nonumber \\
&\vdots& \nonumber
\eea
where the $Z_{pp'}$ are determined sequentially so that
$\la \xi_p' | \xi_p \ra = \delta_{pp'}$. 
We have chosen to take the eigenstate of 
$H_{{\rm low}\,k}$ with lowest energy as 
the starting vector, although any other linear combination 
could have been used. In practice, the modified Gram-Schmidt algorithm 
with re-orthogonalization of Ref.~\cite{gramschmidt} is utilized
to guard against round-off errors. The transformation $Z$ corresponding 
to the Gram-Schmidt hermitization is then given by
\beqn
Z = \sum_{p} | \xi_p \ra \, \la \widetilde{\chi}_p | 
\qquad \text{and} \qquad
Z^{-1} = \sum_{p} | \chi_p \ra \, \la \xi_p | \,.
\eeqn

In contrast to the Gram-Schmidt method, the other hermitization schemes 
are formulated 
in terms of the Lee-Suzuki wave-operator $\omega$, which apparently 
relies on the use of orthogonal projection operators $PQ=0$ 
corresponding to sharp cutoffs. 
In order to generalize the Okubo and Andreozzi hermitization schemes 
to smooth cutoffs, it is necessary to eliminate all references to $\omega$. 
This is easily done
by noting that the bi-orthogonal complement vectors are defined through 
$\la \widetilde{\chi}_p \m \chi_{p'} \ra = \delta_{pp'}$. For a sharp
cutoff, Eq.~(\ref{nonhermmetric}) implies $| \widetilde{\chi}_p
\ra = (P+\omega^{\dagger}\omega) | \chi_p \ra$, and thus
\beqn
P+\omega^{\dagger}\omega = \sum_{p} | \widetilde{\chi}_p \ra \,
\la \widetilde{\chi}_p | \,.
\eeqn 
For smooth cutoffs, the obvious generalization is to construct the operator 
$\sum_{p} | \widetilde{\chi}_p \ra \, \la \widetilde{\chi}_p |$ 
and decompose it as
\beqn
Z^{\dagger}Z= \sum_{p} | \widetilde{\chi}_p \ra \, \la \widetilde{\chi}_p
| \,.
\eeqn
As for sharp cutoffs, the generalized Okubo transformation corresponds to 
$Z=\sqrt{\sum_{p} | \widetilde{\chi}_p \ra \, \la \widetilde{\chi}_p
|}$, where the square root is taken in the eigenbasis of the
positive-definite operator 
$\sum_{p} | \widetilde{\chi}_p \ra \, \la \widetilde{\chi}_p |$.  
Similarly, the Andreozzi hermitization can be obtained from performing 
the appropriate Cholesky decomposition.

\section{Energy-independent RG using smooth cutoffs}
\label{sect:EindepRG}

In this section, we generalize the energy-independent RG equation of
Ref.~\cite{VlowkRG} to smooth cutoffs. For energy-independent interactions, 
the reduced half-on-shell $t$ matrix obeys a Lippmann-Schwinger equation 
with loop integrals smoothly cut off by $f^2(p)$,
\beqn
t(k',k;k^2) = v(k',k) + \frac{2}{\pi} \int_{0}^{\infty} p^2dp \,
\frac{v(k',p) \, f^2(p) \, t(p,k;k^2)}{k^2-p^2} \,,
\eeqn
where the energy-independent low-momentum interaction $\vlowk$ and 
the corresponding half-on-shell $\tlowk$ matrix are given by
\bea
\vlowk(k',k) &=& f(k') \, v(k',k) \, f(k) \,, \\[2mm]
\tlowk(k',k;k^2) &=& f(k') \, t(k',k;k^2) \, f(k) \,.
\eea
We again note that the smooth cutoff regulator is on the loop
momentum but not on external momenta.

Analogous to the derivation for a sharp cutoff~\cite{VlowkRG}, we
impose that the reduced half-on-shell $t$ matrix is independent of the
cutoff, $dt(k',k;k^2)/d\Lambda=0$. 
This choice preserves the on-shell $t$ matrix while also maintaining
energy independence. 
The result is
\begin{multline}
0 = \frac{2}{\pi} \int_{0}^{\infty} p^2dp \, \frac{d v(k',p)}{d\lm} 
\biggl[ \frac{\pi \, \delta(p-k)}{2 k^2} + \frac{f^2(p) \, t(p,k;k^2)}{k^2-p^2}
\biggr] \\[2mm]
- \frac{2}{\pi}\int_{0}^{\infty} p^2dp \, 
\frac{v(k',p) \, \ddlamfsq \, t(p,k;k^2)}{p^2-k^2} \,.
\label{RGE1}
\end{multline}
Next, we express the term in the square brackets
in Eq.~(\ref{RGE1})
by the exact scattering state $| \chi_k \ra$ of $\vlowk$. From
$\vlowk | \chi_k \ra = \vlowk | k \ra + \vlowk \, G_0(k^2) \,
\tlowk | k \ra$, it follows that 
$\la p \m \chi_k \ra = \la p \m k \ra + [ f(p) \, t(p,k;k^2) \, 
f(k) ]/(k^2-p^2)$. We can therefore write Eq.~(\ref{RGE1}) as
\beqn
\frac{2}{\pi} \int_{0}^{\infty} p^2dp \, \frac{d v(k',p)}{d\lm} 
\frac{f(p)}{f(k)} \, \chi_k(p) =
\frac{2}{\pi} \int_{0}^{\infty} p^2dp \: 2 \, \frac{d f(p)}{d\lm} \,
\frac{v(k',p) \, f(p) \, t(p,k;k^2)}{p^2-k^2} \,,
\eeqn
or equivalently
\begin{align}
\frac{2}{\pi} \int_{0}^{\infty} p^2dp \, \frac{d v(k',p)}{d\lm} \,
f(p)\,\chi_k(p) &= \frac{2}{\pi} \int_{0}^{\infty} p^2dp \:
2 \frac{d f(p)}{d\lm} \, \frac{v(k',p)}{p^2-k^2} \, \la p \m \vlowk
\m \chi_k \ra \nonumber \\[4mm]
&= \frac{2}{\pi} \int_{0}^{\infty} p^2dp \:
2\frac{d f(p)}{d\lm} \, v(k',p) \, \la p \m \vlowk \, G(p^2) 
\m \chi_k \ra ,
\end{align}
where $G(E) = (E-H_{{\rm low}\,k})^{-1}$ is the interacting two-nucleon
Green's function. Using the completeness relation, $\int k^2dk \,
\widetilde{\chi}^*_k(p') \, \chi_k(p) = \frac{\pi}{2}\delta(p-p')/p^2$ and
$\vlowk \, G(E) = \tlowk(E) \, G_0(E)$, we have
\beqn
\frac{d}{d\lm} \, v(k',k) = \frac{2}{\pi}\int_{0}^{\infty} p^2dp \, 
\frac{v(k',p) \, \ddlamfsq \, t(p,k;p^2)}{p^2-k^2} \,,
\label{EindepRGE}
\eeqn
which describes the evolution of the reduced low-momentum interaction
with the cutoff.

If we take an energy-independent NN potential as large-cutoff initial 
condition and numerically integrate the RG equation, then the resulting 
$v$ preserves the half-on-shell $\tnn$ matrix for \emph{all} external 
momenta, $t(k',k;k^2)=\tnn(k',k;k^2)$. Therefore, $\vlowk(k',k)=f(k')\,
v(k',k)\,f(k)$ preserves the low-momentum half-on-shell $\tnn$ matrix
up to factors of the smooth cutoff function. In the limit $f(p) \rightarrow 
\theta(\Lambda-p)$ and thus $d[f(p)^2]/d\lm \rightarrow \delta(\Lambda-p)$,
we recover the RG equation for a sharp cutoff~\cite{VlowkRG}
\beqn
\frac{d}{d\Lambda}\vlowk(k',k) = \frac{2}{\pi} \frac{\vlowk(k',\Lambda)
\, \tlowk(\Lambda,k;\Lambda^2)}{1-(k/\Lambda)^2} \,.
\label{sharpRGE}
\eeqn

Finally, we can use the Okubo transformation to hermitize $\vlowk$.
In order to generalize the hermitization to smooth cutoffs, we consider
the sharp-cutoff Okubo transformation under an infinitesimal change 
of the cutoff. In this case, one can show that the RG equation for the
hermitian $\vlowkherm$ is given by a symmetrized version of 
Eq.~(\ref{sharpRGE}),
\begin{multline}
\frac{d}{d\Lambda}\vlowkherm(k',k) = \frac{1}{\pi} \biggl[
\frac{\vlowkherm(k',\Lambda) \,
\tlowkherm(\Lambda,k;\Lambda^2)}{1-(k/\Lambda)^2} \\[2mm]
+\frac{\tlowkherm(k',\Lambda;\Lambda^2) \,
\vlowkherm(\Lambda,k)}{1-(k'/\Lambda)^2} \biggr] \,.
\end{multline}
A simple generalization of the Okubo transformation to smooth cutoffs
is therefore obtained by symmetrizing the smooth-cutoff RG equation, 
Eq.~(\ref{EindepRGE}), to obtain
\begin{multline}
\frac{d}{d\lm} \, \overline{v}(k',k) =
\frac{1}{\pi} \int_{0}^{\infty} p^2dp \,
\biggl[\frac{\overline{v}(k',p) \ddlamfsq \, 
\overline{t}(p,k;p^2)}{p^2-k^2} \\[2mm]
+ \frac{\overline{t}(k',p;p^2) \, \ddlamfsq \, 
\overline{v}(p,k)}{p^2-k'^2} \biggr] \,.
\end{multline}
As in the previous section, the hermitian low-momentum interaction 
$\vlowkherm$ preserves the low-momentum fully-on-shell $\tnn$ matrix, up 
to factors of the regulator function $\tlowk(k,k;k^2)=f^2(k)\,
\tnn(k,k;k^2)$, and the deuteron binding energy.
The freedom in the hermitization method can be expressed
as an auxiliary condition $d t(k',k;k^2)/d\Lambda = (k^2-k'^2) \,
\Phi(k',k)$, where $\Phi(k',k)$ is a function with $\lim_{k \to k'} 
(k^2-k'^2) \, \Phi(k',k) = 0$. The above RG equation derived from the 
Okubo transformation makes a particular choice for $\Phi(k',k)$. 

The numerical solution of 
the energy-independent RG equation is complicated by the $t$ matrix 
calculation involved in each step. 
The computational overhead slows down the ODE solver significantly.
In addition, the RG equation involves two-dimensional
interpolations and principal-value integrals over narrowly
peaked functions. Therefore, it is easy to introduce small errors 
at each step that can accumulate as the cutoff is lowered. Finally, 
some potential models exhibit spurious resonances (at order GeV 
energies and momenta)~\cite{Kai} and these need to be subtracted
before solving the energy-independent RG equation for these potentials.
Because of these difficulties, we have exclusively used the three-step method
from Sect.~\ref{sect:EdepRG} to generate the results presented in the next
section.


\section{Results}
\label{sect:2b}

In this section, we apply the formalism discussed in 
Sect.~\ref{sect:EdepRG} to derive hermitian, low-momentum interactions 
with smooth cutoffs. Electromagnetic contributions are included
in the evolution to low momenta.
Since all of the following results are for the hermitian $\vlowk$,
we drop the overbar hereafter. 
In selecting a smooth regulator function satisfying the conditions
of Eq.~(\ref{eq:regulator}), there is obviously much freedom,
which parallels the freedom of field redefinitions in low-energy
effective theories and the functional RG~\cite{Latorre:2000jp}.
However, there are trade-offs in the choice.
A smoother cutoff will dampen more the artifacts of a theta-function
regulator but will distort more the phase shifts for momenta near the cutoff.

We present results for two choices for $f(k)$, each with a range of
parameters.  
These are the exponential form used in current
chiral EFT potentials~\cite{N3LO,N3LOEGM} with integer $n$ determining
the smoothness,
\beqn
f(k) = e^{-(k^2/\lm^2)^n} \,,
\label{eq:expreg}
\eeqn
and a Fermi-Dirac form with a sharper cutoff achieved
with smaller $\epsilon$,
\beqn
f(k) = \frac{1}{1+e^{(k^2 - \lm^2)/\epsilon^2}} \,.
\label{eq:fdreg}
\eeqn
It may be interesting in the future
to explore other choices for the regulator
function, but the general features and advantages are covered by
the above choices. Other possible regulators (used in
different applications, e.g., see Ref.~\cite{Liao2000}) 
include a power law form with integer $n$,
\beqn
f(k) = \frac{1}{1 + (k^2/\Lambda^2)^n} \,,
\label{eq:powreg}
\eeqn
a hyperbolic tangent form with an $\epsilon$ parameter that
plays a similar role as in the Fermi-Dirac function,
\beqn
f(k) = \frac{1}{2} \left[ 1 + \tanh\left(
\frac{\Lambda^2-k^2}{\Lambda k \epsilon}\right) \right] \,,
\label{eq:tanhreg}
\eeqn
a complementary error function with $\epsilon$ parameter, 
\beqn
f(k) = \frac{1}{2} \, 
{\rm erfc}\left(\frac{k-\Lambda}{\epsilon}\right)\,,
\label{eq:erfcreg}
\eeqn
and a Strutinsky averaging with $\epsilon$ parameter,
\beqn
f(k) = \frac{1}{2}\left[
{\rm erf}
\left(\frac{\Lambda^2+k^2}{\epsilon^2} \right) +
{\rm erf}
\left(\frac{\Lambda^2-k^2}{\epsilon^2} \right)
\right] \,.
\label{eq:erfreg}
\eeqn
In each case, the function and its derivatives are continuous,
and the parameter $n$ or $\epsilon$ controls the smoothness.
  
\begin{figure}[t]
\begin{center}
\includegraphics*[width=3.5in]{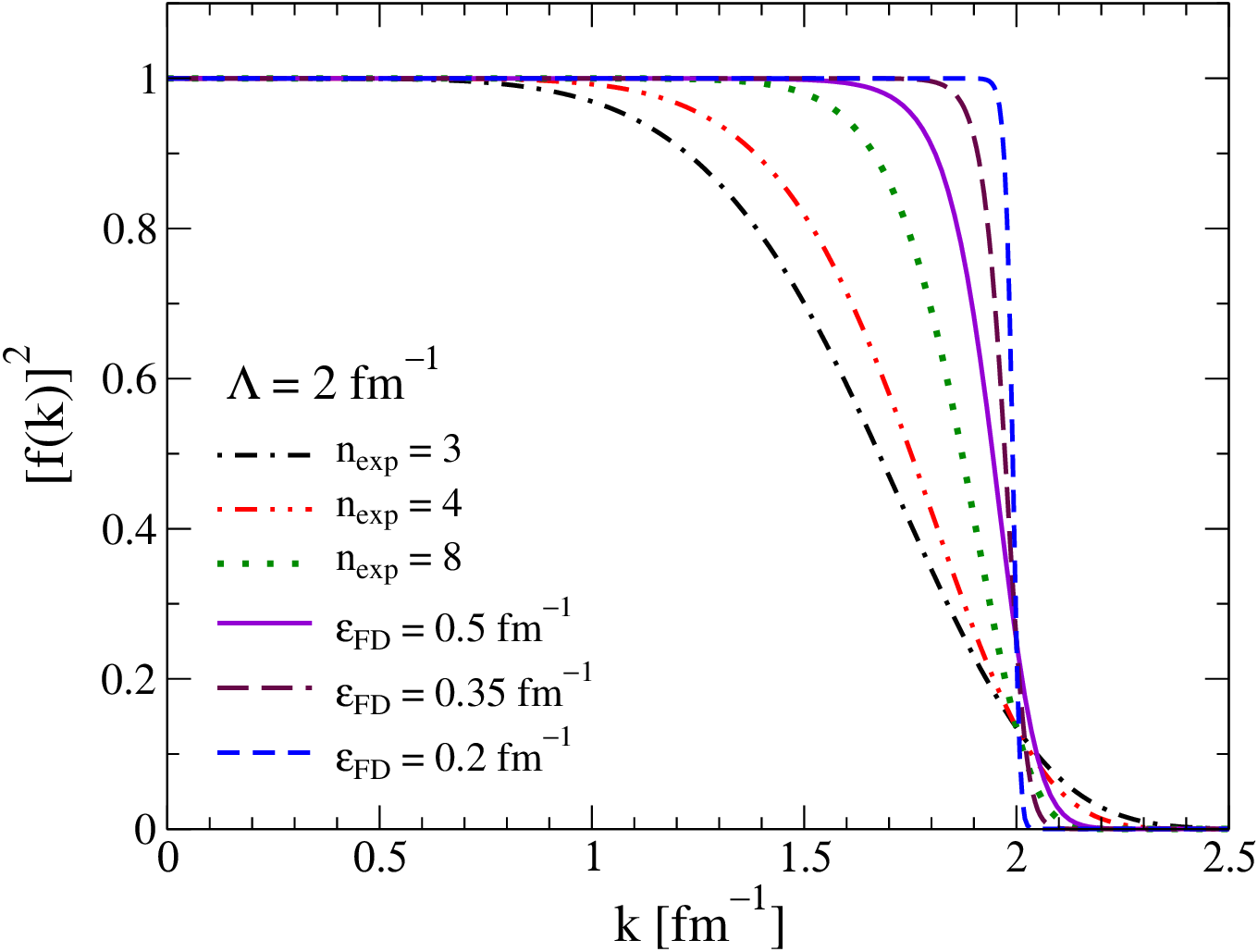}
\end{center}
\caption{Plots of the exponential and Fermi-Dirac
regulators squared as a function of momentum $k$ for
$\Lambda = 2\,\mbox{fm}^{-1}$ and a range of 
parameters $n$ and $\epsilon$.}
\label{fig:regulators1}
\end{figure}

In Fig.~\ref{fig:regulators1}, the quantity $[f(k)]^2$ is plotted against $k$
for fixed cutoff $\Lambda = 2\,\mbox{fm}^{-1}$ using some candidate
parameterizations of the exponential and Fermi-Dirac forms.  
While a sharp cutoff (for which $f(k) = \theta(\Lambda - k)$) 
preserves the on-shell
$T$ matrix up to the cutoff, the $T$ matrix for a smooth cutoff 
is multiplied by $[f(k)]^2$, leading to distortions in the phase shifts
near the cutoff.
The exponential regulator
(``exp'') from Eq.~(\ref{eq:expreg}) with $n = 3$ corresponds to what is
used in N$^3$LO chiral potentials~\cite{N3LO,N3LOEGM}; 
it is evident that this regulator applied in the present context
will significantly distort the phase shifts for momenta
well below $\Lambda$.
As $n$ is increased, the regulator gets sharper; for numerical reasons,
$n=10$ is probably the practical upper limit.  
The Fermi-Dirac form (``FD'') 
interpolates smoothly between sharp and smooth
as a function of the $\epsilon$ parameter. 
It causes less distortion for lower momenta than the exponential regulator.

\begin{figure}[t]
\begin{center}
\includegraphics*[width=2.7in]{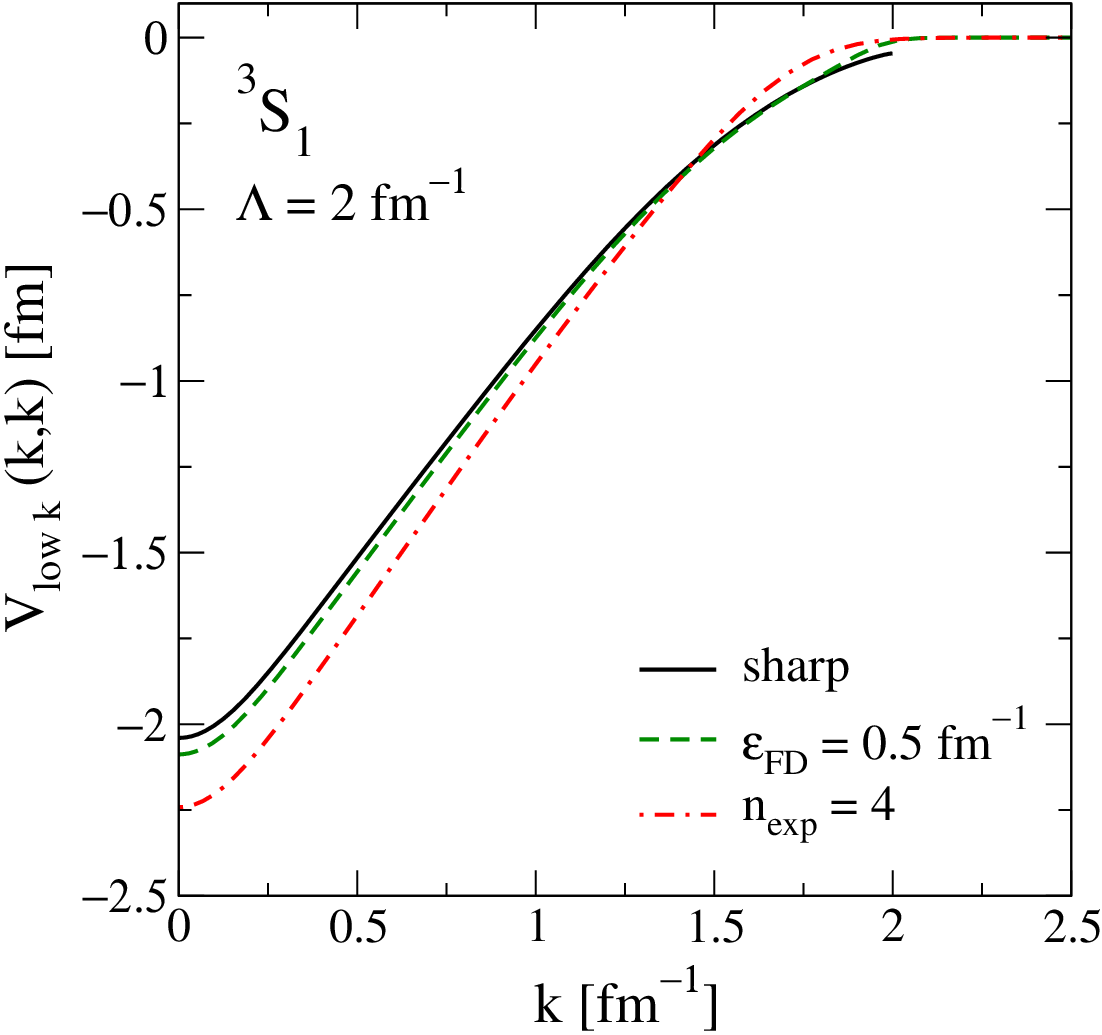}
\hfill
\includegraphics*[width=2.7in]{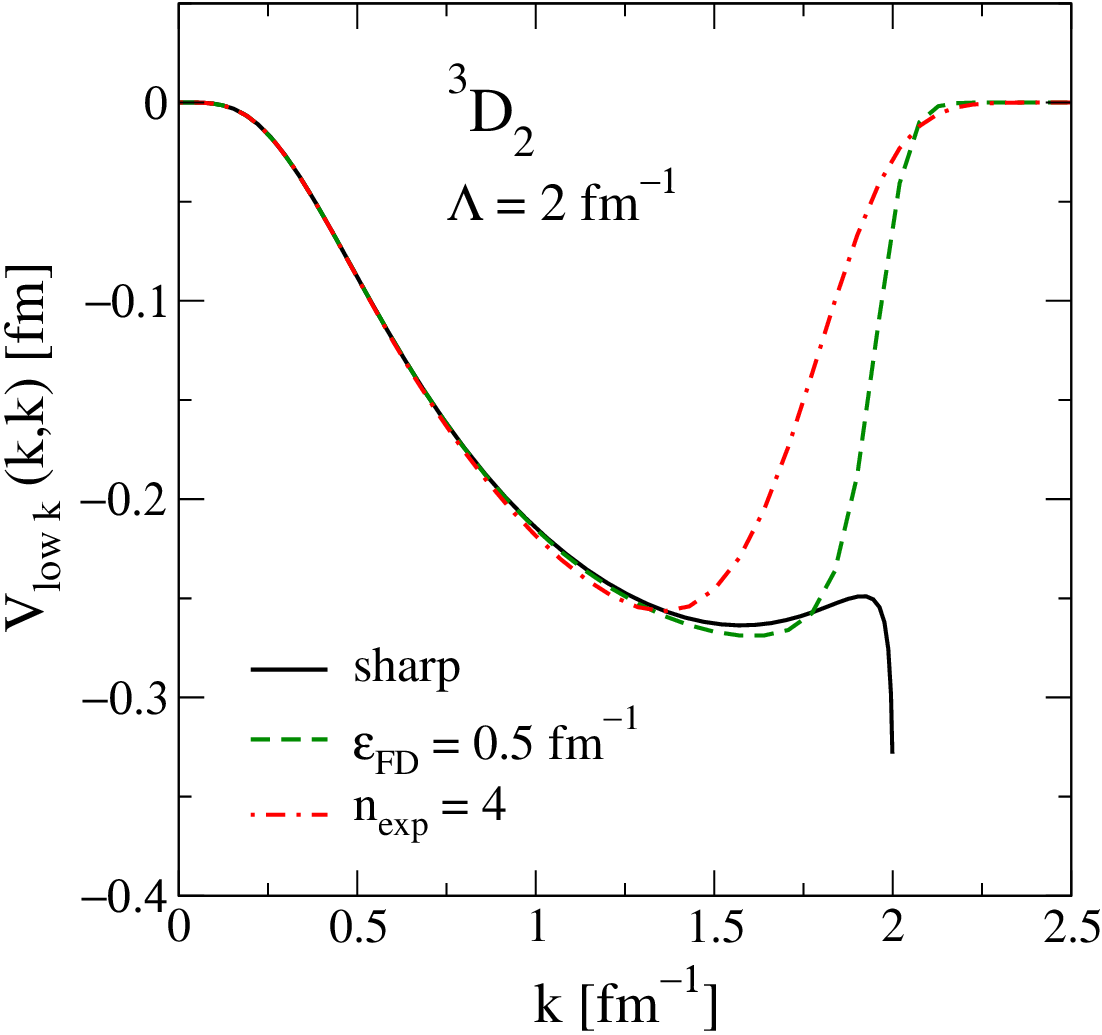}
\end{center}
\caption{Diagonal matrix elements $\vlowk(k,k)$ for $\Lambda 
= 2\,\mbox{fm}^{-1}$, derived from the N$^3$LO chiral potential 
of Ref.~\cite{N3LO} with a sharp and two smooth regulators.}
\label{fig:vlowksmooth}
\end{figure}

The effects of different 
regulators on $\vlowk$ potentials with the same starting
(``bare'') potential and the same cutoff are illustrated in
Fig.~\ref{fig:vlowksmooth}.
In channels where the potential is close to zero at $\Lambda$,
such as the $^3$S$_1$ partial wave, the differences between 
sharp and smooth are slight,
particularly as the regulator gets sharper ($n_{\rm exp}$ is smoother
than $\epsilon_{\rm FD} = 0.5\,\mbox{fm}^{-1}$; see 
Fig.~\ref{fig:regulators1}). However, the difference
in other channels can be striking, as seen in Fig.~\ref{fig:vlowksmooth}
for the $^3$D$_2$ partial wave. The observed cusp-like behavior is 
due to reproducing the phase shifts for momenta up to the cutoff.
The existence of sharp regulator artifacts is not a problem in principle,
as the potential is not an observable, but in practice it can lead
to convergence problems at the 10--100\,keV level in the deuteron and
triton.

\begin{figure}[t]
\begin{center}
\includegraphics*[width=4.2in]{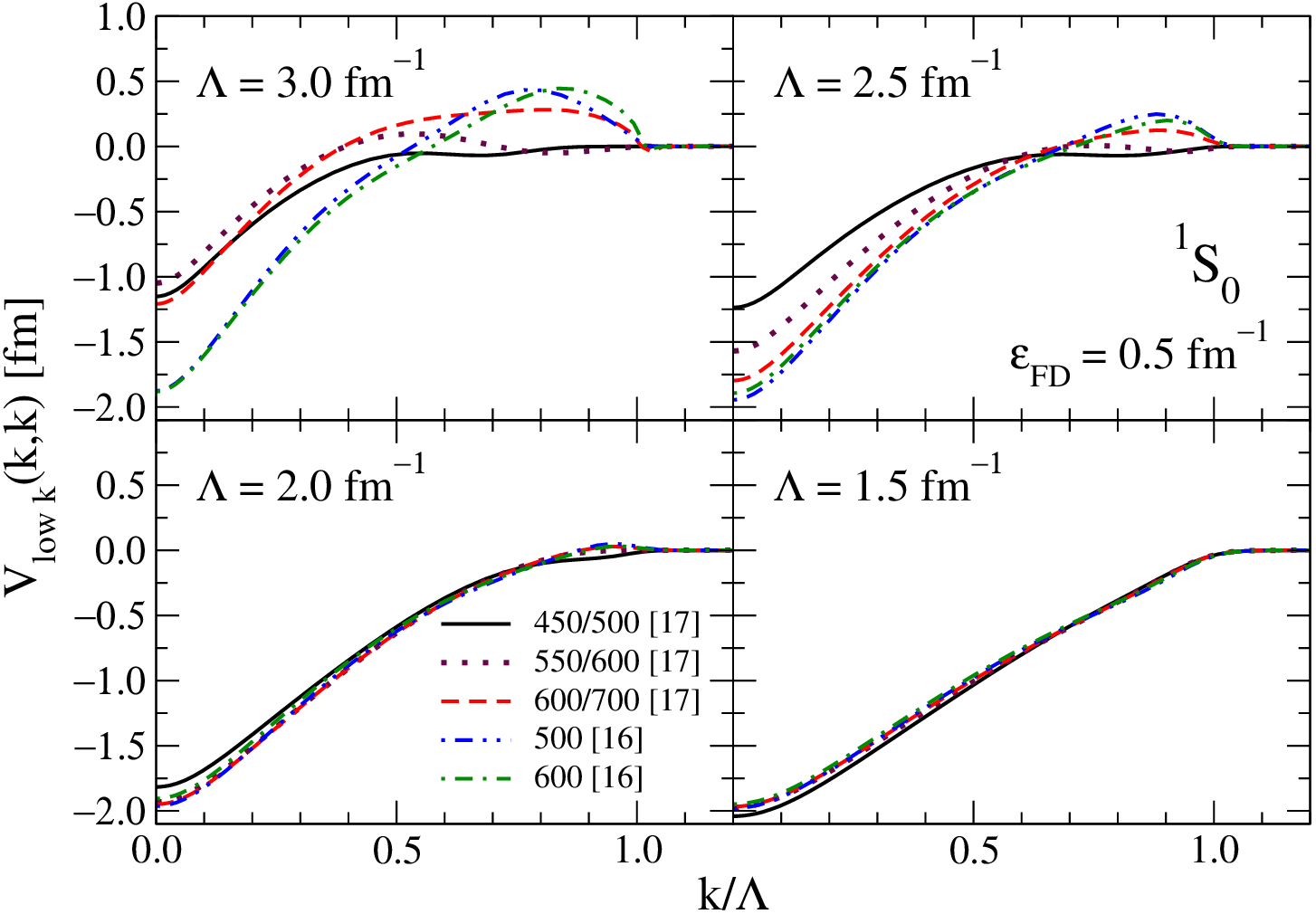}
\end{center}
\vspace*{.2in}
\begin{center}
\includegraphics*[width=4.2in]{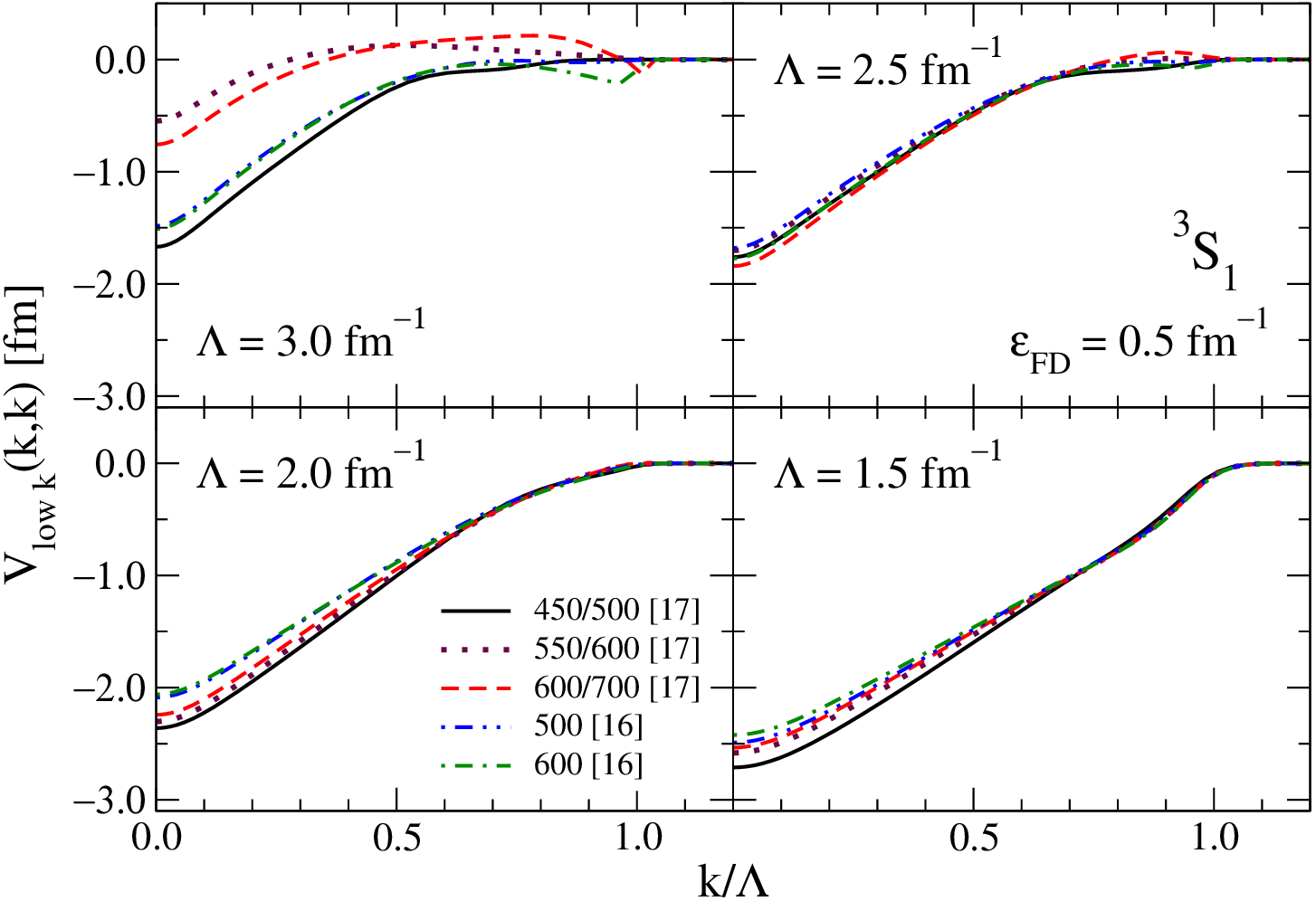}
\end{center}
\caption{The ``collapse'' of $\vlowk$ interactions derived
from N$^3$LO chiral potentials with a Fermi-Dirac regulator
($\epsilon_{\rm FD} = 0.5\,\mbox{fm}^{-1}$)
as the cutoff is lowered in the $^1$S$_0$ and $^3$S$_1$ channels.
The diagonal matrix elements $\vlowk(k,k)$ are shown, but as in
Ref.~\cite{Vlowk2}, we find similar results for the off-diagonal
matrix elements.
The different lines correspond to different starting potentials
with the corresponding cutoffs, $\Lambda$~\cite{N3LO} or $\Lambda/
\widetilde{\Lambda}$~\cite{N3LOEGM}, in MeV given in the legends.
For Ref.~\cite{N3LOEGM}, $\widetilde{\Lambda}$ is the spectral
function cutoff.}
\label{fig:vlowkcollapsen3lo}
\end{figure}

One of the striking properties of $\vlowk$ with a sharp cutoff is 
the ``collapse'' of different high-precision NN potentials to almost
the same low-momentum potential as the cutoff
is lowered below $3\,\mbox{fm}^{-1}$~\cite{Vlowk2}.  
This behavior is expected as a consequence of the same long-range
pion-exchange interaction together with phase shift
equivalence of the potentials up to $k \approx 2.1 \fmi$.
Therefore, it is not surprising that it remains a property of
$\vlowk$ interactions derived
with smooth regulators, as shown
in Fig.~\ref{fig:vlowkcollapsen3lo} for a set of chiral potentials with
different initial (``bare'') cutoffs.
Note that the low-momentum interactions from different starting
potentials are close but \emph{not identical}.
The differences will be paralleled by differences in the corresponding 
short-range three-body interactions.

\begin{figure}[t]
\begin{center}
\includegraphics*[width=4.8in]{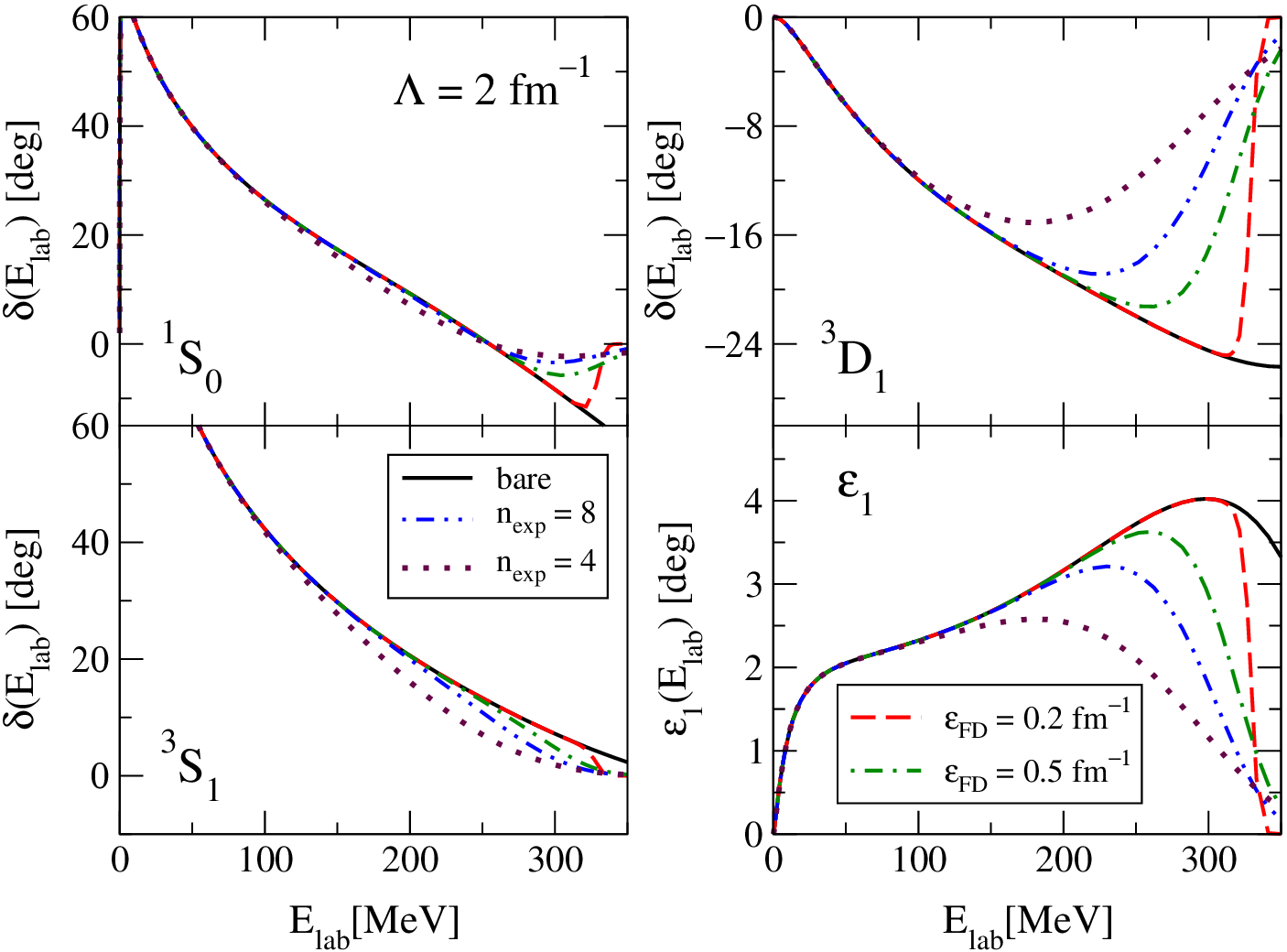}
\end{center}
\caption{The effects of different regulators on the initial
phase shifts $\delta$ and the mixing parameter
$\varepsilon_1$ as a function
of laboratory energy $E_{\rm lab}$ is shown for selected channels.
The phase shifts are calculated using the 
N$^3$LO chiral potential from Ref.~\cite{N3LO}.}
\label{fig:errorlim}
\end{figure}

Low-momentum interactions with smooth cutoffs reproduce the initial
phase shifts up to factors of the regulator function.
The error in phase shifts 
\emph{due to the regulator alone} is illustrated in
Fig.~\ref{fig:errorlim} for some representative two-body phase shifts
as a function of laboratory energy.
In particular, for each energy, the on-shell T-matrix from a bare potential
(in this case the N$^3$LO chiral potential from Ref.~\cite{N3LO})
is multiplied
by $[f(k)]^2$ using the momentum $k$ corresponding to that energy. 
This corresponds to the distortion that would be present if numerical
errors in constructing $\vlowk$ were negligible.
The latter are documented next and are small.
We have no universal rule for deciding whether a distortion is acceptable;
it depends on how it propagates to the observable in question.
For example,
for the low-energy bound-state of the deuteron, none of the
distortions in Fig.~\ref{fig:errorlim} is important.
The distortion is analogous to the error band from a chiral EFT truncation
(but we have not formulated a corresponding power counting rule).
Therefore, we expect that there is no concern if the distortion is 
comparable to the EFT truncation error.
In addition, for low-energy properties, the error incurred here can 
be absorbed by the short-range part of the corresponding three-body 
interactions. Consequently, the propagation of phase-shift errors to 
many-body observables can only be studied after including three-nucleon
forces.
It is evident, however, that $n_{\rm exp} = 8$ and $\epsilon_{\rm FD} = 0.5$
only distort minimally. We will show below that these are also good 
choices for convergence.
In future applications, the regulator effects
can be tested by varying the parameters determining the smoothness.

\begin{figure}[t]
\begin{center}
\includegraphics*[width=2.7in]{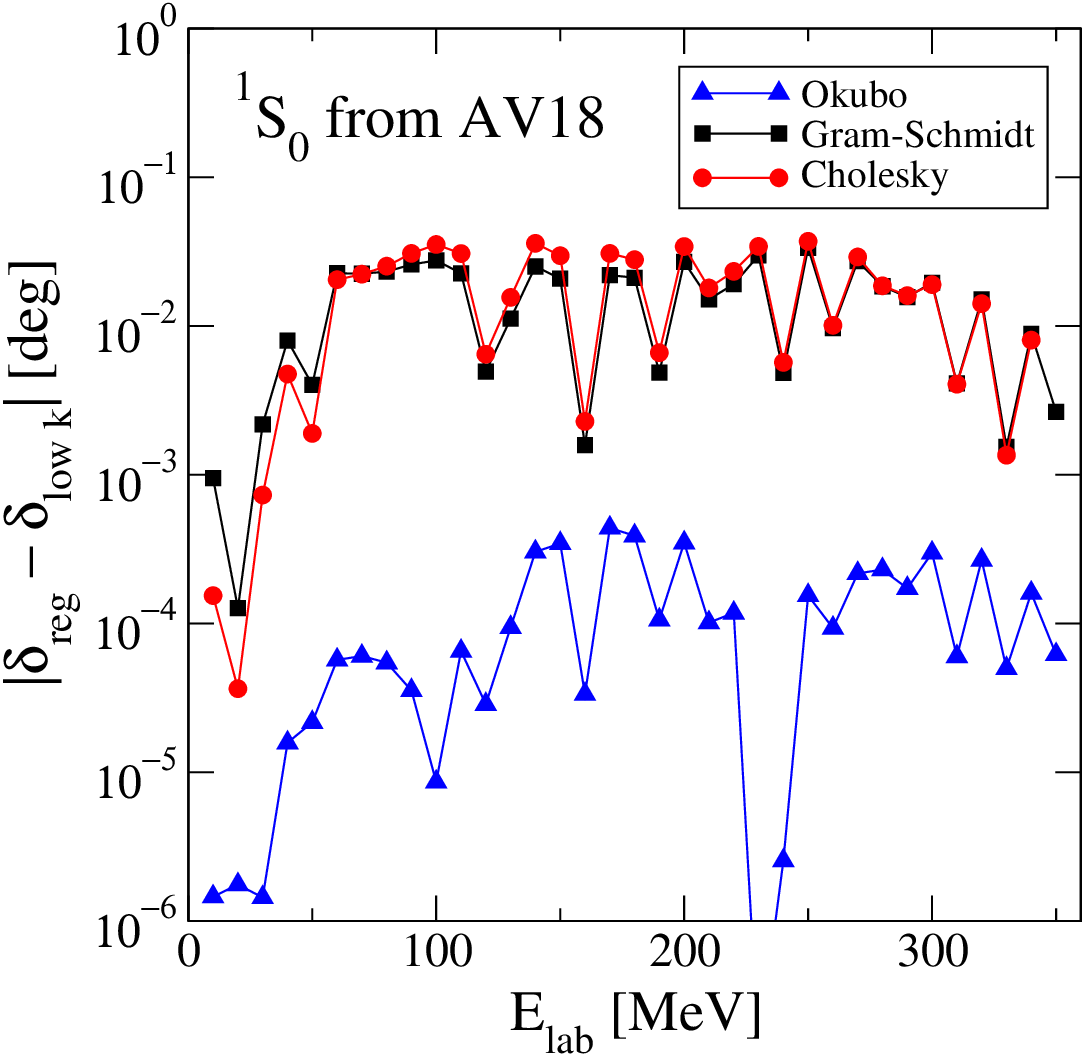}
\hfill
\includegraphics*[width=2.7in]{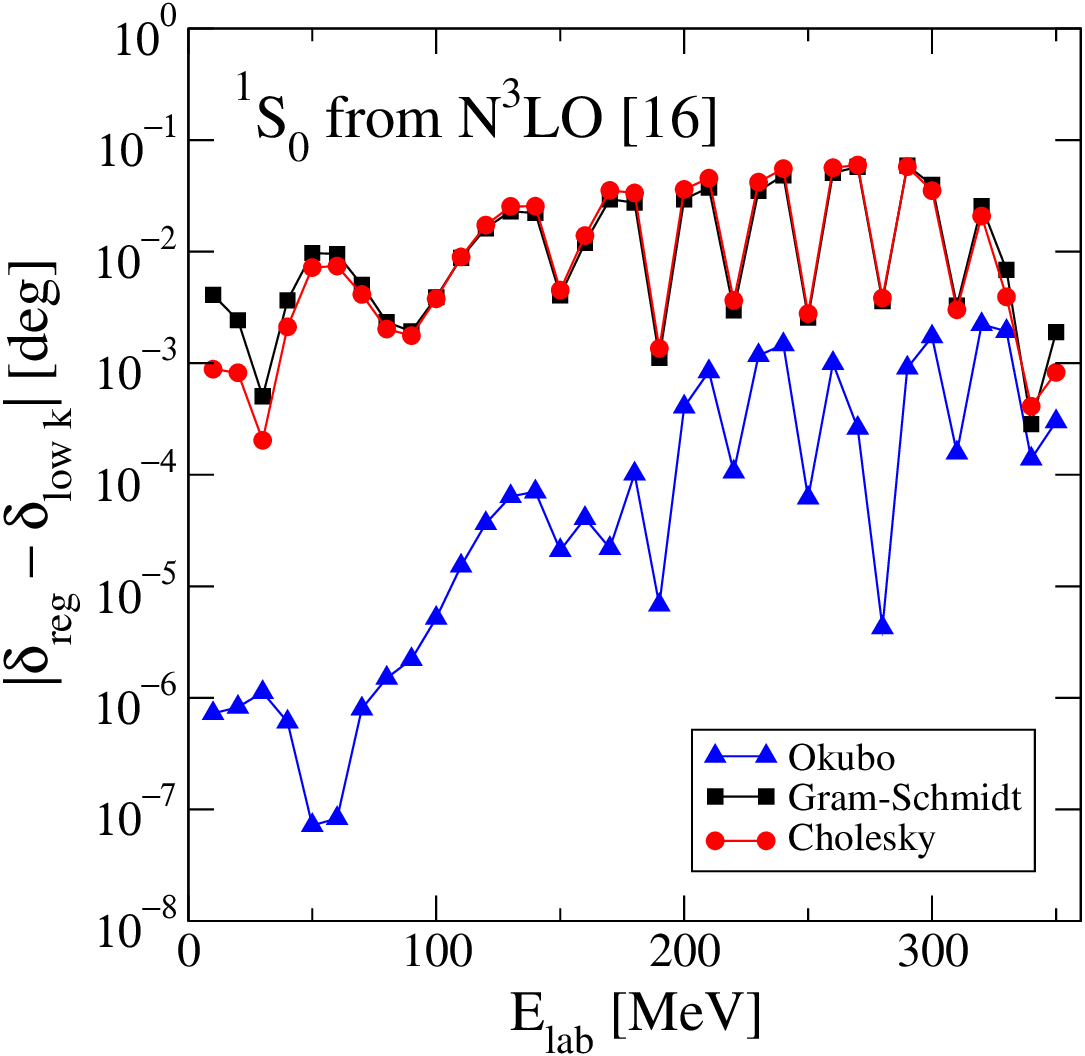}
\end{center}
\caption{Errors in phase shifts relative to \emph{regulated} 
bare phase shifts
for three hermitization methods using the exponential regulator
with $\Lambda = 2\,\mbox{fm}^{-1}$.
The potentials used on the left were derived from Argonne 
$v_{18}$~\cite{AV18} with $n_{\rm exp} = 4$
and those used on the right were derived from the 
N$^3$LO chiral potential from Ref.~\cite{N3LO} with $n_{\rm exp} = 8$.}
\label{fig:errornum}
\end{figure}

Phase shifts can also differ from those calculated from the input
(``bare'') potential because of numerical errors in generating the
low-momentum interaction.
In Fig.~\ref{fig:errornum}, the effects of numerical errors on the
phase shifts for different hermitization methods (but fixed
regulator) are isolated
by plotting the difference of the calculated low-momentum 
phase shifts and the \emph{distorted} bare phase shifts.
We conclude that the Okubo hermitization is numerically
more robust than the Gram-Schmidt or Cholesky methods,
with errors in the phase shifts of order $10^{-4}$ degrees below 200\,MeV
and then varying up to $10^{-3}$ degrees depending on the regulator
and the initial potential.
Consequently, we use the Okubo hermitization in the following
unless otherwise specified.
For fixed hermitization but different
regulators, we have found that relatively smooth cutoffs
(e.g., $\epsilon_{\rm FD} = 0.5\,\mbox{fm}^{-1}$) achieve 
$10^{-4}$ degree accuracy with a moderate (of order 50) Gauss points,
but sharper cutoffs (e.g., $\epsilon_{\rm FD} = 0.2\,\mbox{fm}^{-1}$)
have errors for energies above $E_{\rm lab} = 200\,\mbox{MeV}$ that
can grow as large as $10^{-1}$ degrees. 
Greater accuracy can be obtained with a more carefully prescribed
distribution of points.\footnote{We note that 
the present scheme occasionally and unsystematically 
leads to numerical ``glitches'' in the low-momentum interaction, 
particularly for cutoffs close to the
bare cutoff.  These are manifested as discontinuities in the potential
and are signaled by large discrepancies in calculated matrix elements.
We have found that adjusting the momentum grid removes the glitches, 
but we do not yet have a preventive fix.} 

\begin{figure}[t]
\begin{center}
\includegraphics*[width=2.7in]{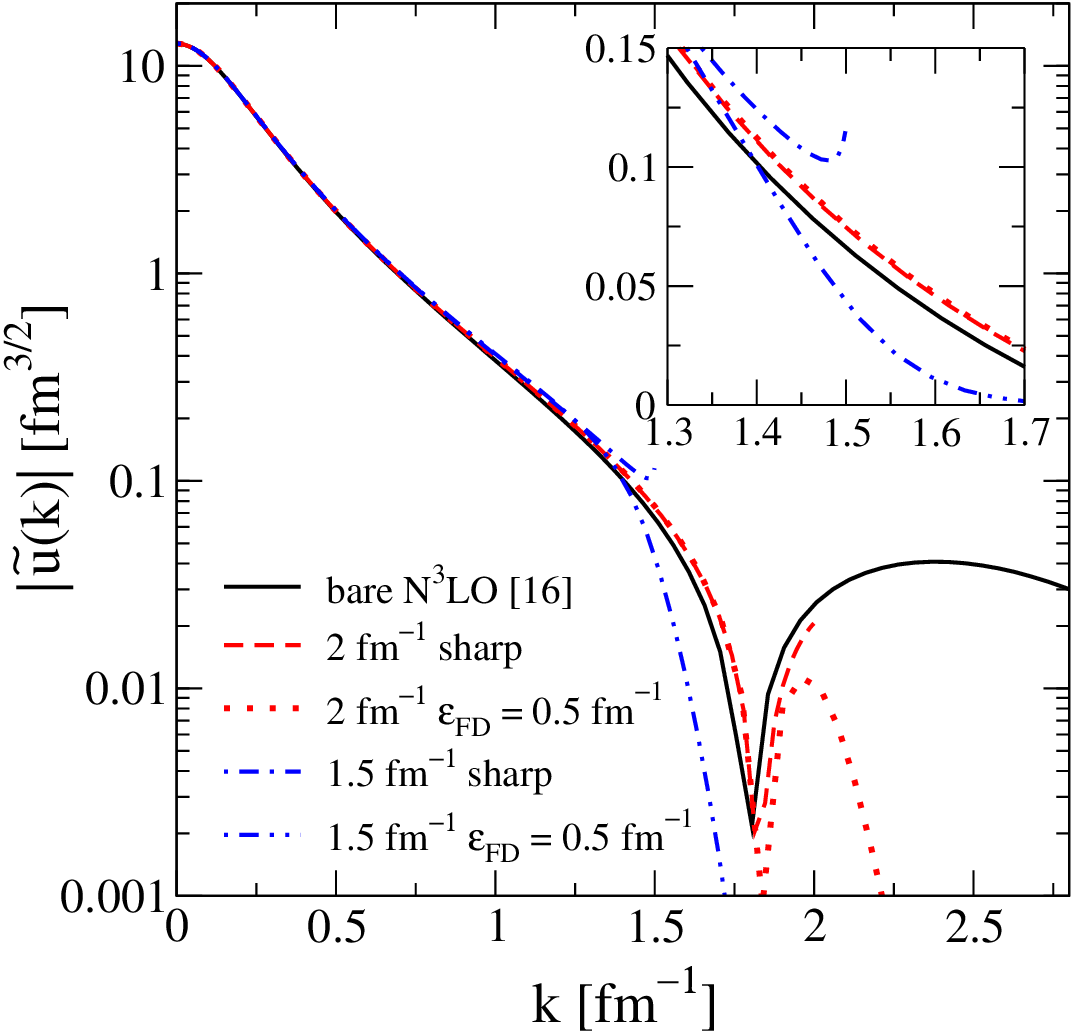}
\hfill
\includegraphics*[width=2.7in]{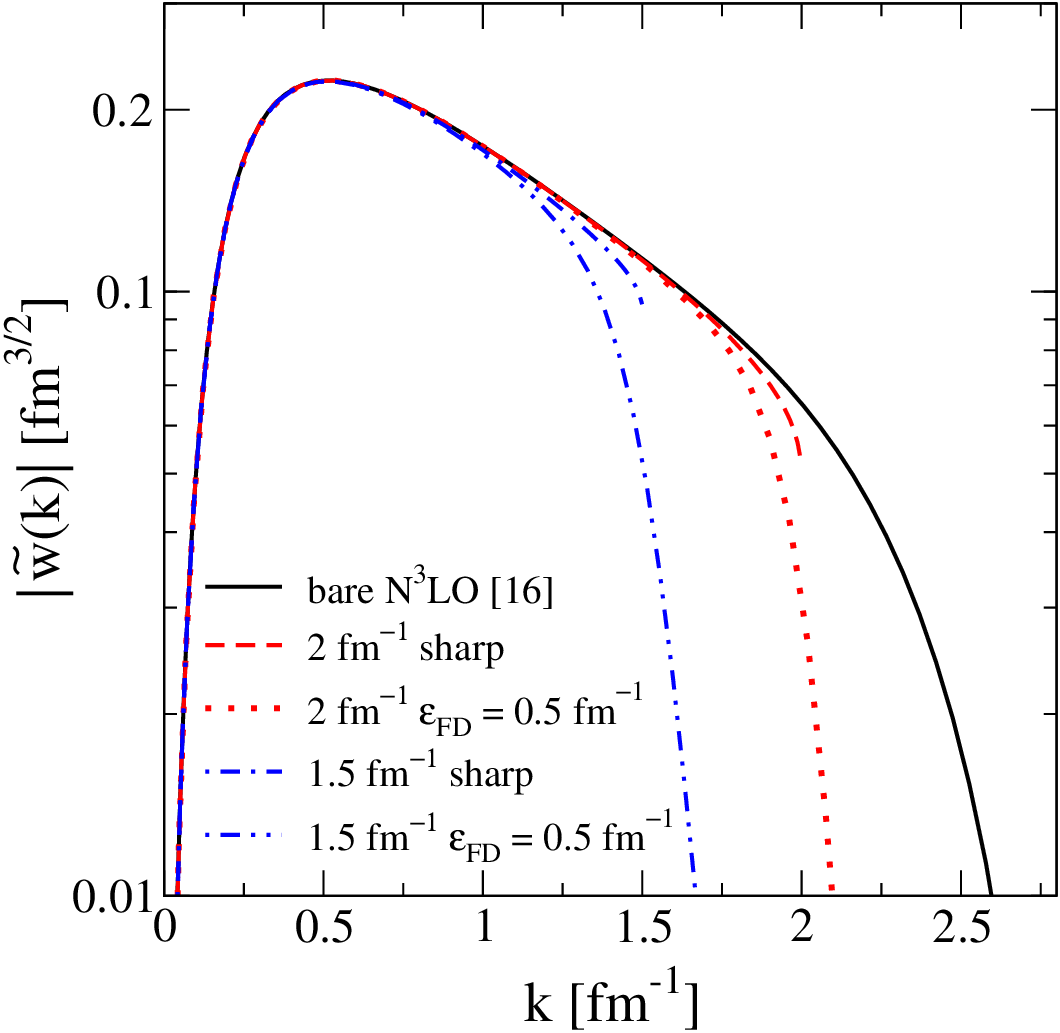}
\end{center}
\caption{S--state and D--state components of the deuteron wave function in 
momentum space ($\widetilde{u}(k)$ and $\widetilde{w}(k)$ respectively) 
for the bare N$^3$LO chiral potential from Ref.~\cite{N3LO} and those 
derived using smooth and sharp cutoffs at $\Lambda = 2.0\,\mbox{fm}^{-1}$
and $1.5\,\mbox{fm}^{-1}$.}
\label{fig:deuteronwfs}
\end{figure}

\begin{figure}[t]
\begin{center}
\includegraphics*[width=2.7in]{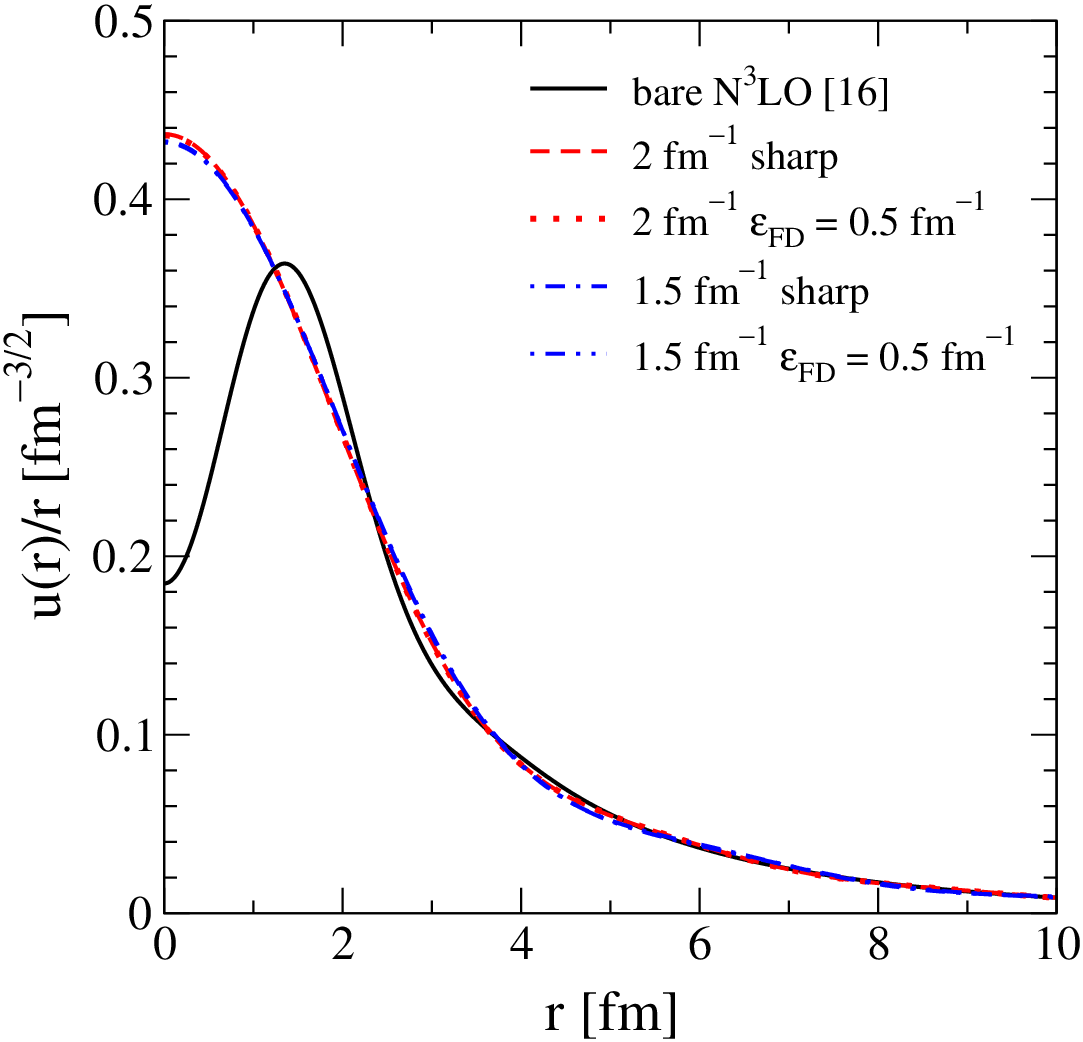}
\hfill
\includegraphics*[width=2.7in]{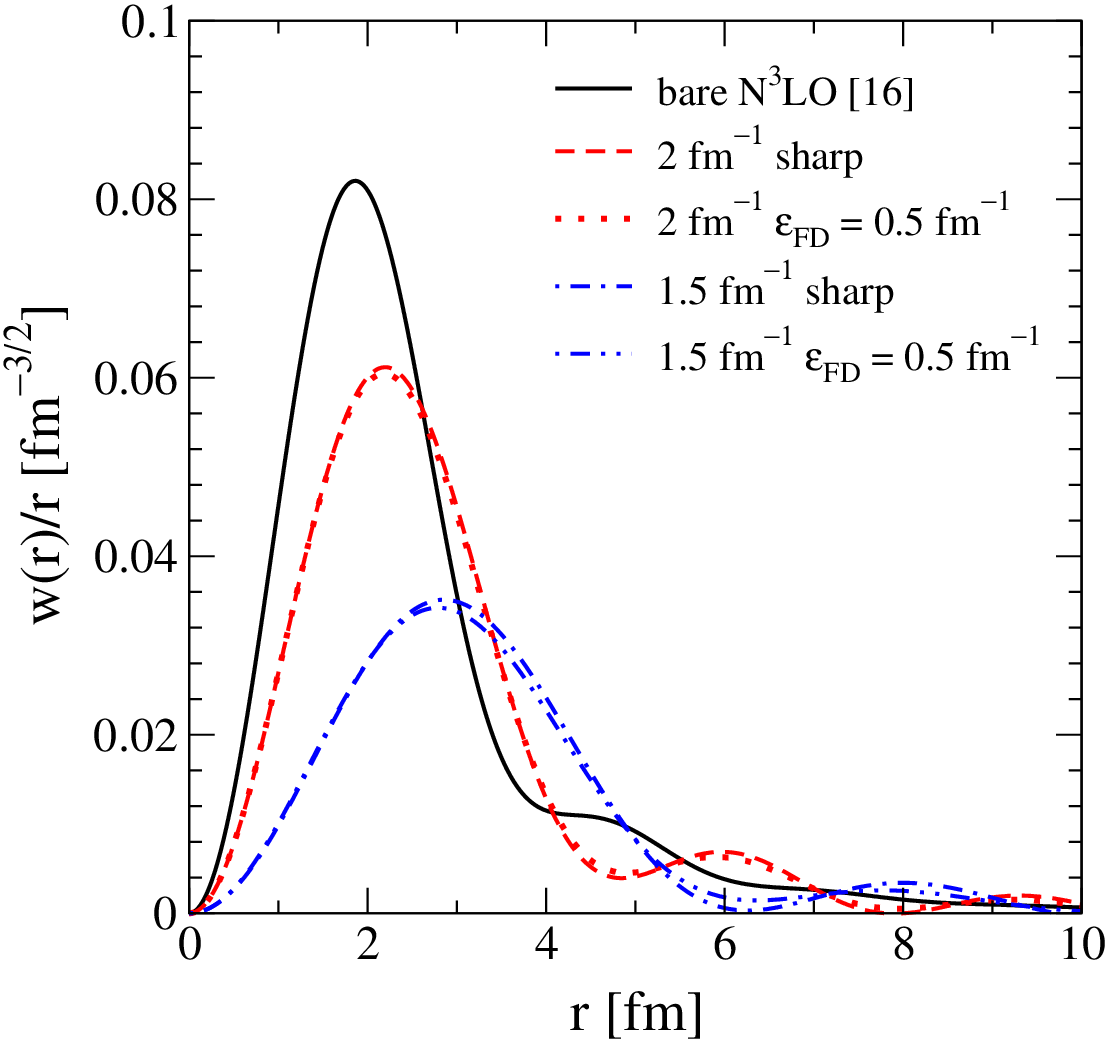}
\end{center}
\caption{Deuteron wave functions in coordinate space for smooth and
sharp cutoffs as in Fig.~\ref{fig:deuteronwfs}.}
\label{fig:deuteronwfsp}
\end{figure}

The deuteron wave functions for smooth and sharp
cutoffs are contrasted in momentum space in
Fig.~\ref{fig:deuteronwfs} and in
coordinate space in
Fig.~\ref{fig:deuteronwfsp}.
We follow the notation of Ref.~\cite{Epelbaum:1999dj}, with S--wave
and D--wave components denoted in coordinate space
by $u$ and $w$ respectively, and with tildes in momentum space.
They are normalized as
$\int_0^\infty dr\, [u(r)^2 + w(r)^2] = 1$ and
$\int_0^\infty dk\,k^2\, [\wt u(k)^2 + \wt w(k)^2] = 1$.
The sharp-cutoff wave functions develop cusp-like structure
in momentum space below $2\,\mbox{fm}^{-1}$ (see inset in 
Fig.~\ref{fig:deuteronwfs}), which are removed
by the Fermi-Dirac (or any other smooth) regulator.
The different momentum-space behavior is evident in the coordinate-space
wave functions as smaller amplitudes in the large-distance oscillations
with increasing smoothing (for the S--state component this is visible
only under additional magnification).
Note that the ``wound'' in the N$^3$LO coordinate-space
wave function is removed by
running down the cutoff (we plot $u(r)/r$ to make the suppression near
the origin more explicit).
This feature of the wave functions leads to more perturbative behavior
in nuclear matter~\cite{Bogner_nucmatt} as well as in few-body 
systems~\cite{Bogner:2006tw,Bogner:2006a}.

\begin{figure}[t]
\begin{center}
\includegraphics*[width=4.2in]{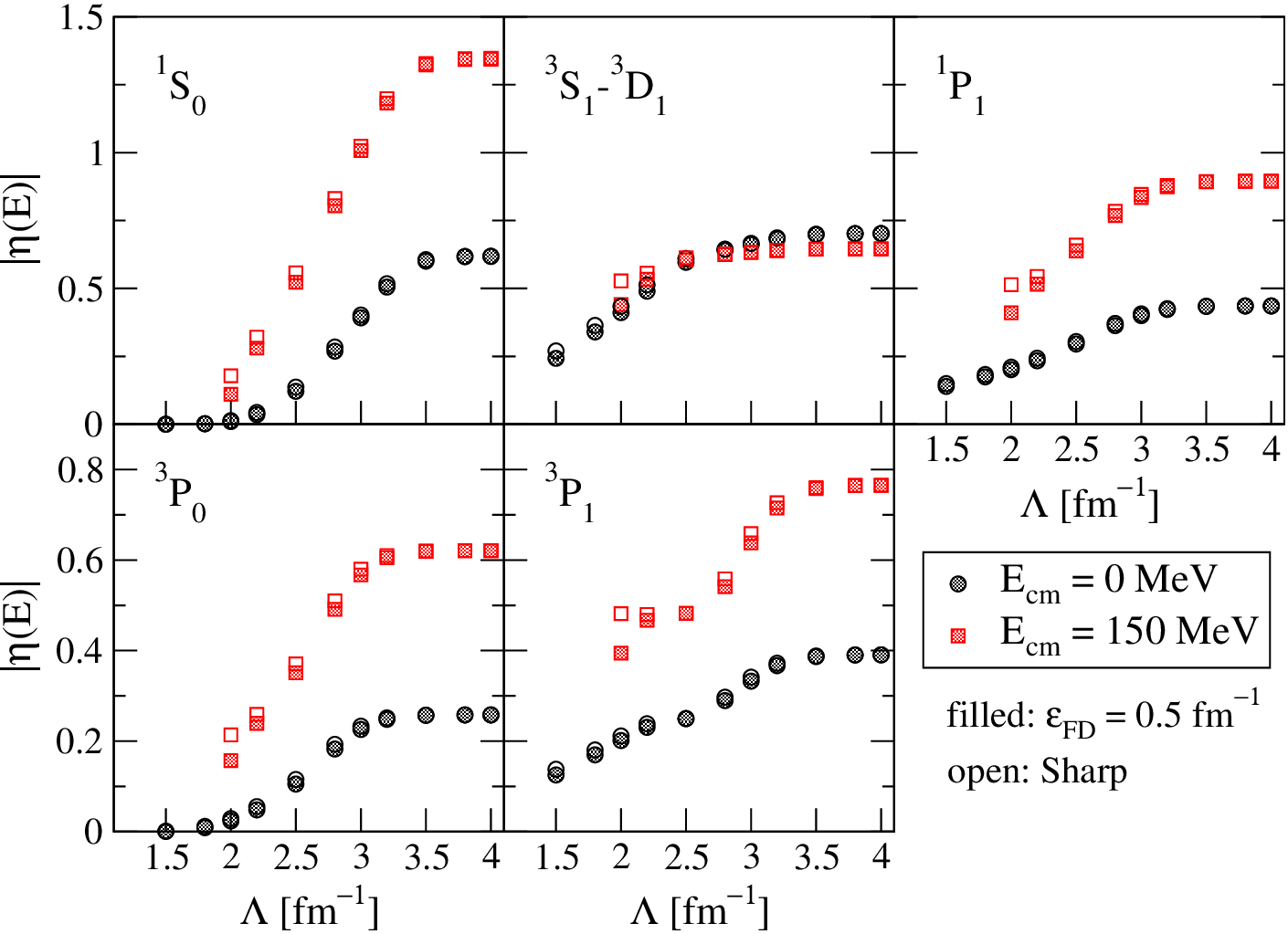}
\end{center}
\caption{The largest repulsive Weinberg eigenvalues at two energies as
a function of cutoff for selected partial waves. The $\vlowk$ interactions
are derived from the N$^3$LO chiral potential of Ref.~\cite{N3LO}.}
\label{fig:weinberg}
\end{figure}

The ``perturbativeness'' of $\vlowk$ interactions with sharp cutoffs
was examined in Ref.~\cite{Bogner:2006tw} using Weinberg eigenvalues 
as a diagnostic. For details on the Weinberg analysis we refer the
reader to Refs.~\cite{Bogner:2006tw,Weinberg}.
In Fig.~\ref{fig:weinberg}, we show the largest repulsive Weinberg 
eigenvalues as a function of the cutoff for selected channels, using
the N$^3$LO chiral potential from Ref.~\cite{N3LO}, which is constructed
with a cutoff of 500\,MeV.
Although this is already a fairly soft potential, we still
observe the characteristic decrease with low-momentum cutoff
starting as high as $3.5\,\mbox{fm}^{-1}$ (rather than at
$2.5\,\mbox{fm}^{-1}$, as one might naively expect). This
translates into weaker correlations in many-body wave functions
(and therefore better convergence). The rate of decrease is largely 
independent of the smoothness of the cutoff. 

\begin{figure}[t]
\begin{center}
\includegraphics*[width=2.7in]{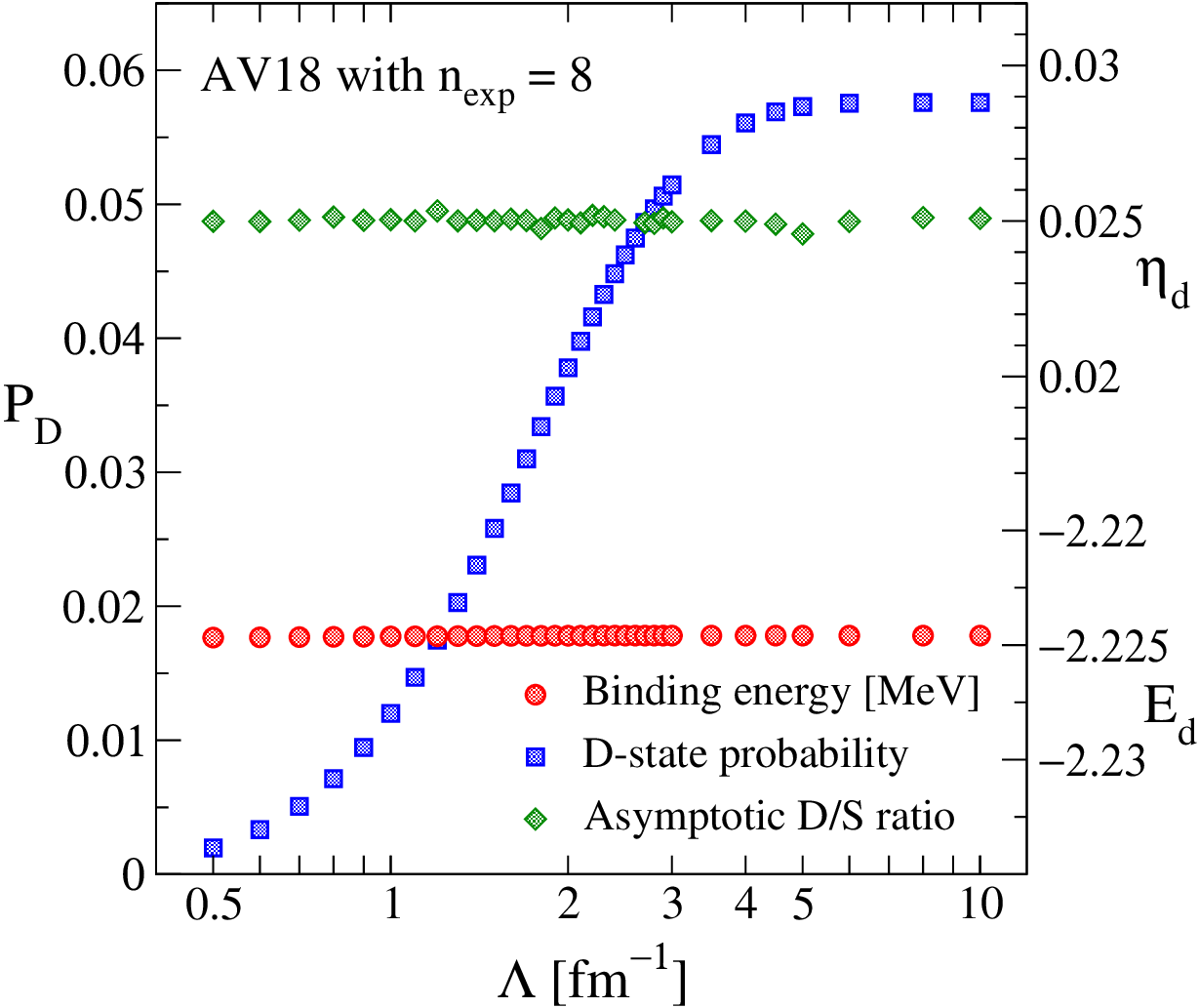}
\hfill
\includegraphics*[width=2.7in]{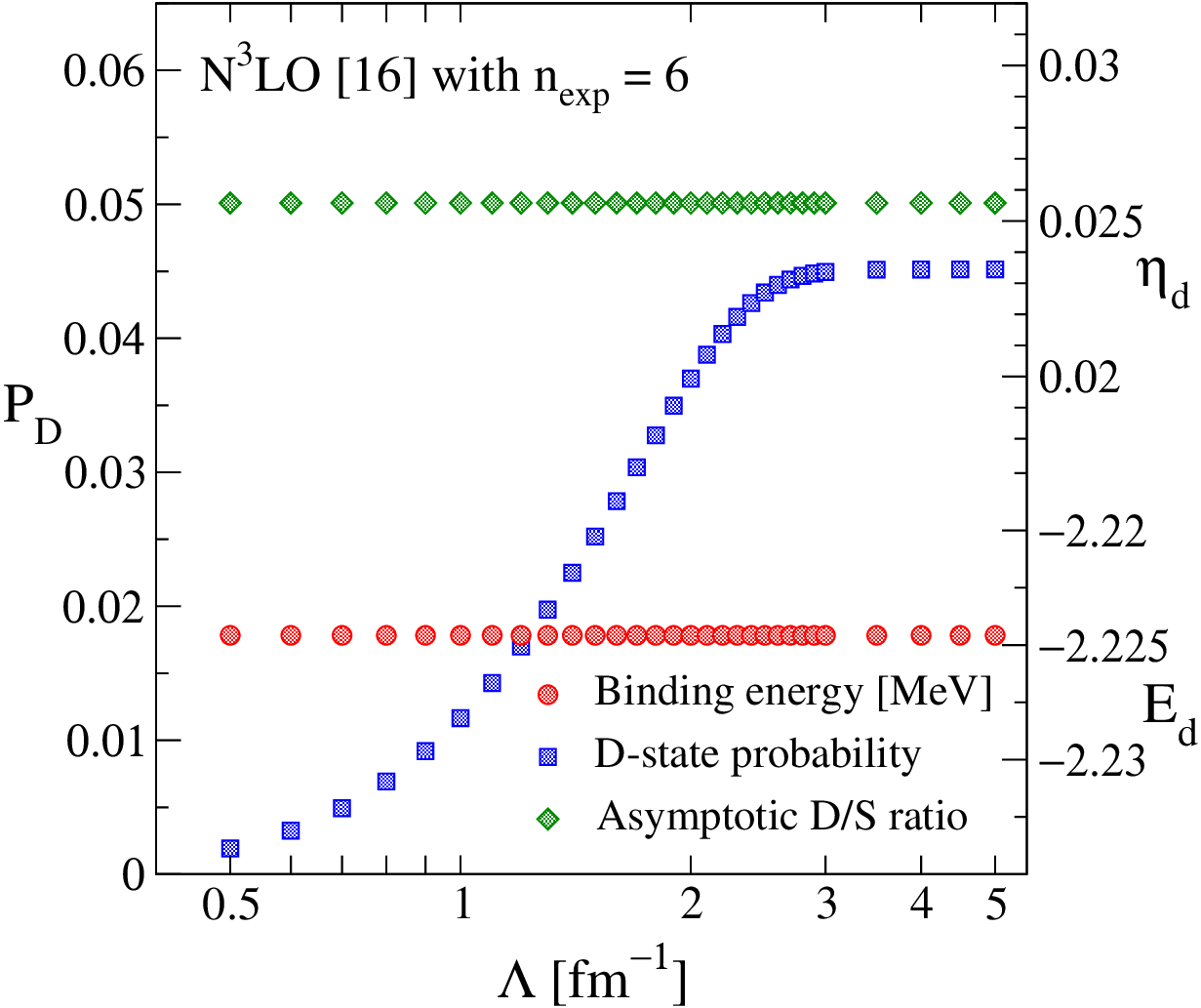}
\end{center}
\caption{D--state probability $P_{\rm D}$ (left axis), binding energy 
$E_d$ (lower right axis), and asymptotic D/S ratio $\eta_{d}$
(upper right axis) of the deuteron as a function of the cutoff, starting 
from the Argonne $v_{18}$~\cite{AV18} 
(left) and the N$^3$LO chiral potential of
Ref.~\cite{N3LO} (right) with different smooth regulators.}
\label{fig:properties}
\end{figure}

Various deuteron properties are shown in Fig.~\ref{fig:properties}
as a function of the cutoff using potentials derived from the Argonne
$v_{18}$ potential~\cite{AV18} 
with exponential regulator $n_{\rm exp}=8$ on the left,
and from the N$^3$LO chiral potential of Ref.~\cite{N3LO} with
exponential regulator $n_{\rm exp}=6$ on the right.
Plotted on the left axis is the D--state probability $P_D$, defined as
\beqn
P_D \equiv \int_0^\infty\! dr \, w(r)^2
= \int_0^\infty\! dk\, k^2\, \wt w(k)^2 \,.
\label{eq:PD}
\eeqn
The cutoff dependence reflects the fact that $P_{\rm D}$ is not an 
observable~\cite{Amado,Friar}. The D--state probability evolves
with the short-range part of the potential and, in particular, the 
short-range tensor interaction decreases as the cutoff is lowered.
This decrease is desirable to reduce correlations in 
many-body wave functions~\cite{Bogner_nucmatt}.
The qualitative change in $P_{\rm D}$ is the same for other
regulators and hermitization schemes (the Gram-Schmidt procedure
is used in Fig.~\ref{fig:properties}).

In contrast to the D--state probability, the deuteron binding energy 
$E_d$ and the asymptotic D/S ratio $\eta_{d}$ are observables
and thus cutoff independent (up to numerical tolerances), as shown
by the right axes in Fig.~\ref{fig:properties}. The asymptotic D/S ratio 
is calculated here by an extrapolation of the ratio of the deuteron
wave function D-- to S--state components to the deuteron pole at 
$k^2 = -m E_d$~\cite{Amado},
rather than the conventional approach of extrapolating the 
$^3$S$_1$--$^3$D$_1$ mixing angle $\varepsilon_1$.
That is, we evaluate $-\widetilde{w}(k^2)/\widetilde{u}(k^2)$ on the
Gauss mesh for positive $k^2$, and then make a (near-linear) 
extrapolation to $k^2 = -m E_d$. The constancy of $\eta_{d}$
in Fig.~\ref{fig:properties}
directly refutes the claim of Ref.~\cite{Nakamura:2005uw} that this
quantity cannot be reproduced using low-momentum interactions.

\begin{figure}[t]
\begin{center}
\includegraphics*[width=2.7in]{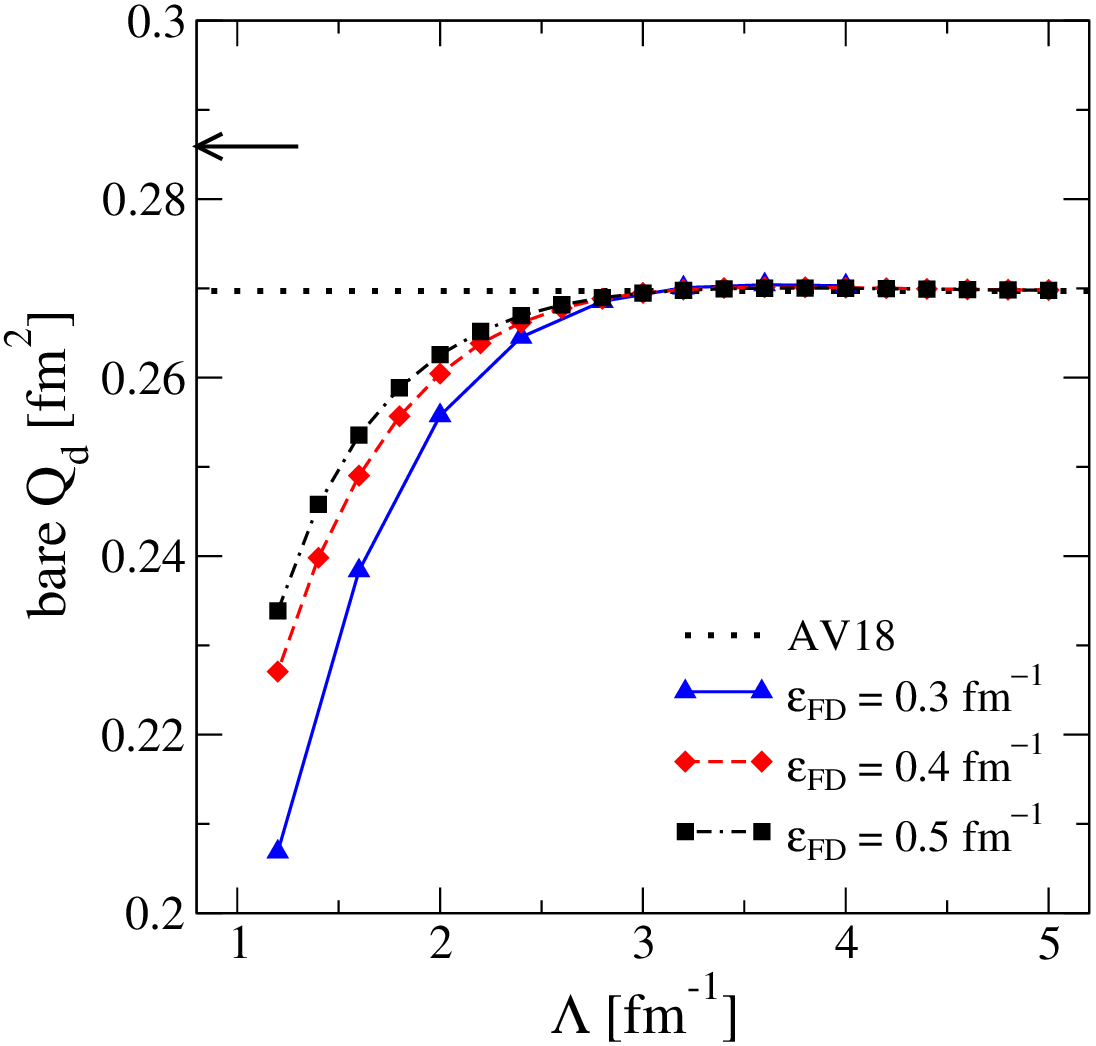}
\hfill
\includegraphics*[width=2.7in]{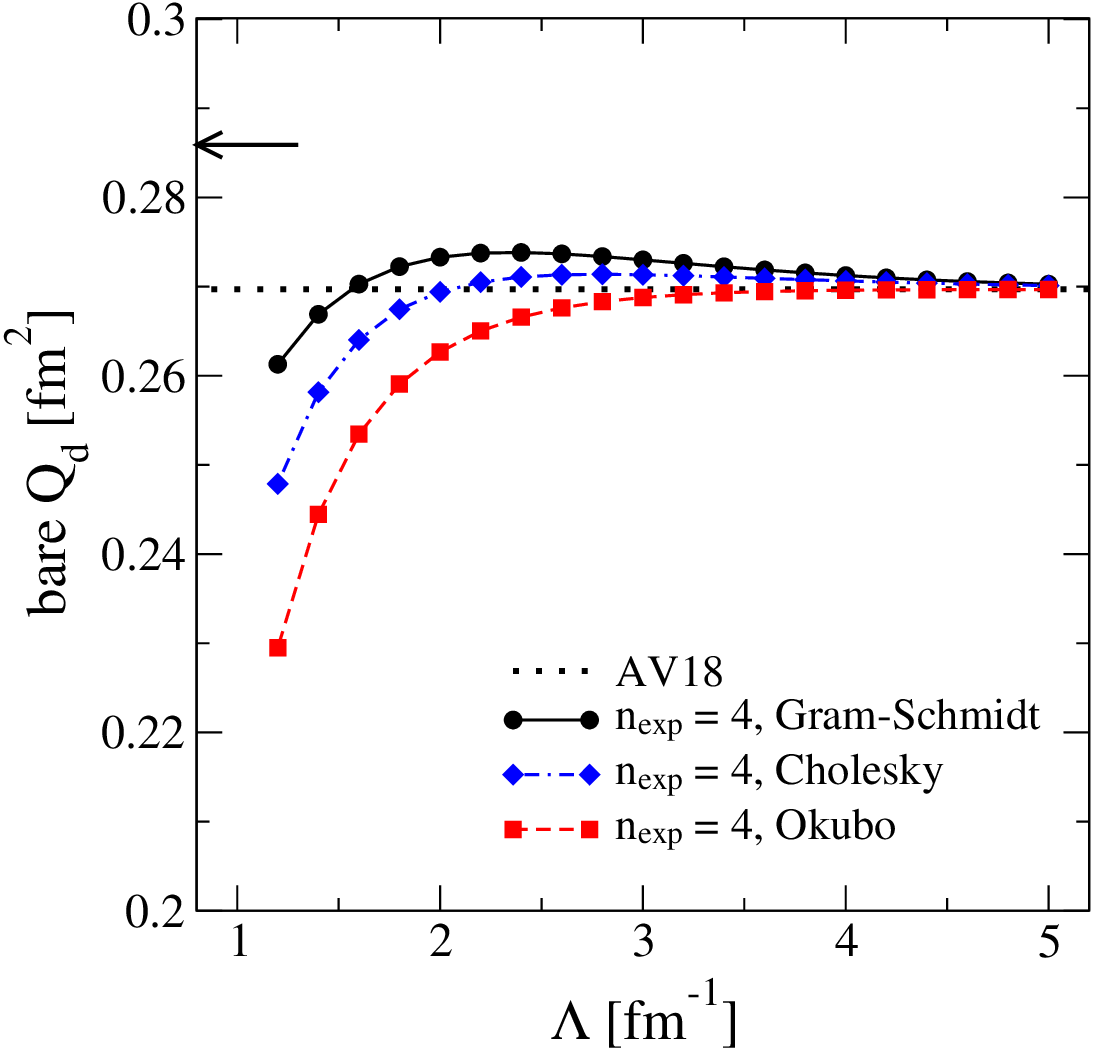}
\end{center}
\caption{The quadrupole moment of the deuteron $Q_d$ calculated with the
bare operator as a function of the cutoff for different smoothness
regulators (left) and for different hermitization schemes (right).  
The low-momentum interactions are derived from the Argonne $v_{18}$ 
potential~\cite{AV18},
and the experimental quadrupole moment is indicated with an arrow.}
\label{fig:Qd}
\end{figure}

\begin{figure}[t]
\begin{center}
\includegraphics*[width=2.7in]{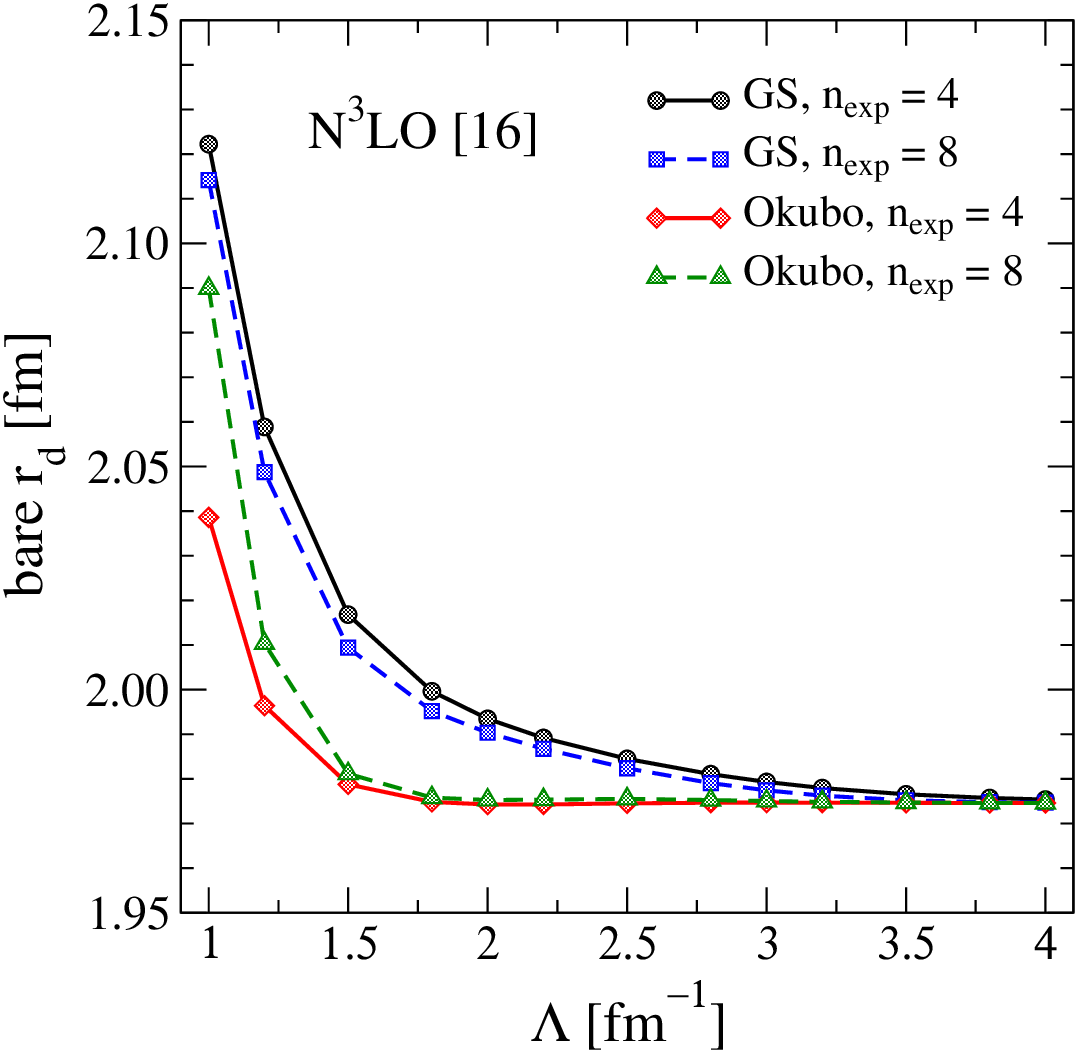}
\hfill
\includegraphics*[width=2.7in]{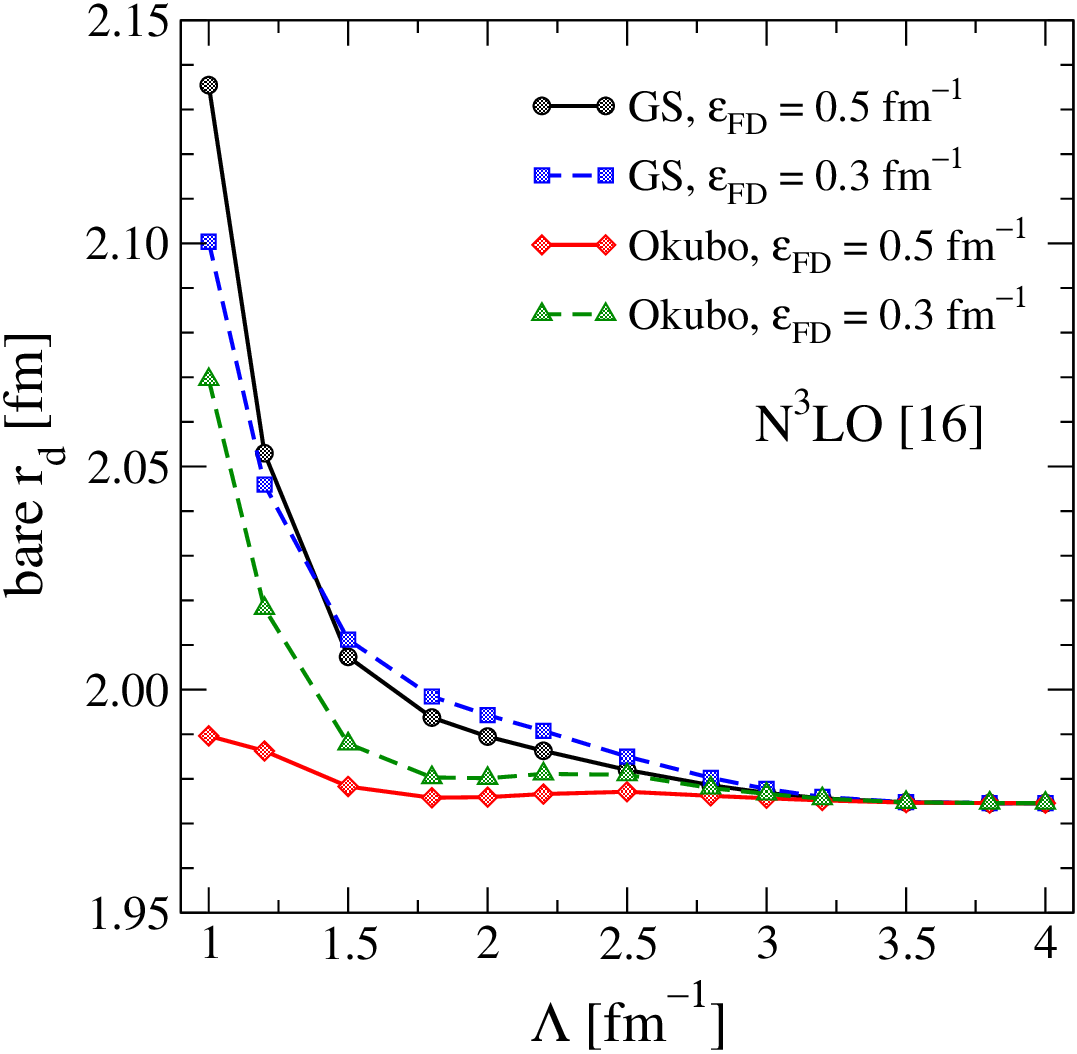}
\end{center}
\caption{The rms radius $r_d$ of the deuteron
calculated with the bare operator as a function of the cutoff 
for different regulators and hermitization schemes.  
The low-momentum interactions are derived using the N$^3$LO
chiral potential from Ref.~\cite{N3LO}.}
\label{fig:rsq}
\end{figure}

\begin{figure}[t]
\begin{center}
\includegraphics*[width=2.7in]{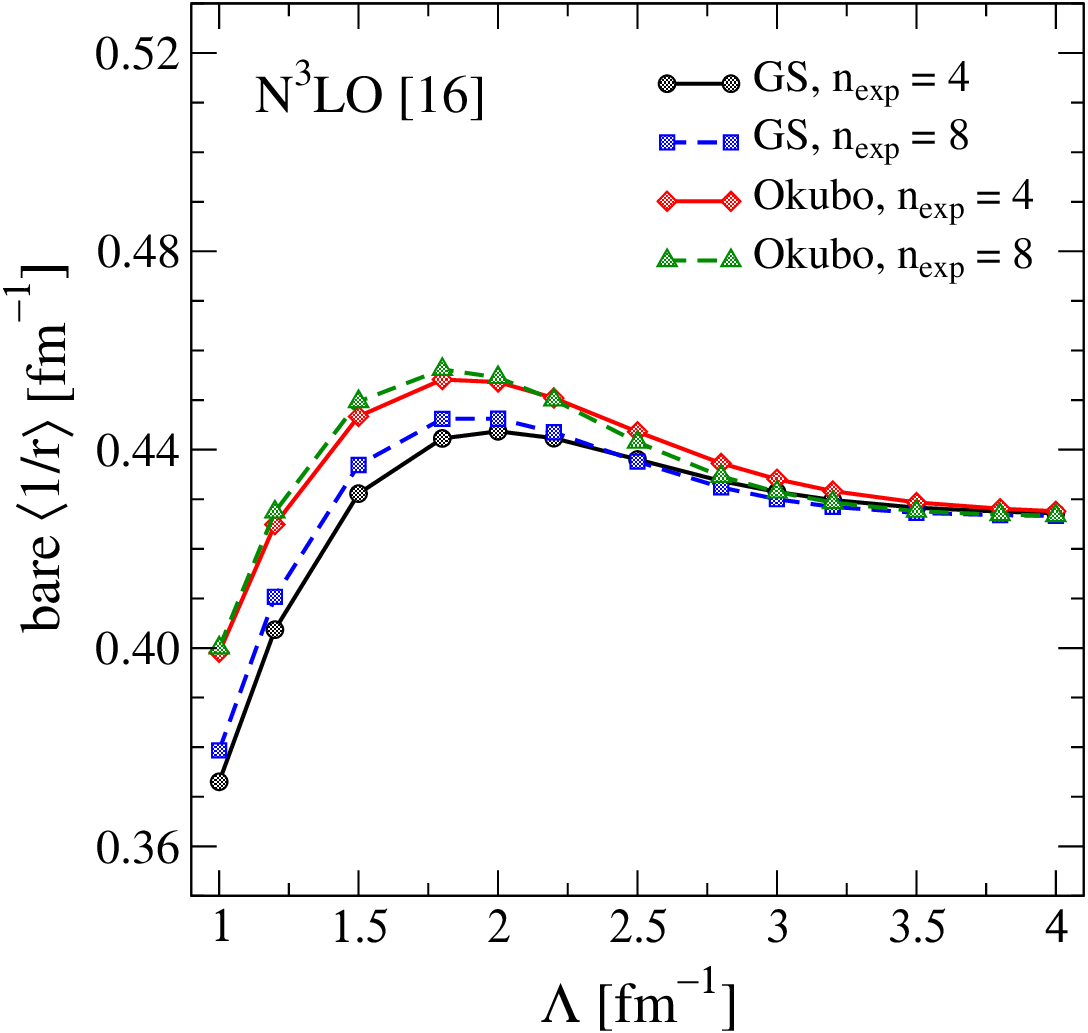}
\hfill
\includegraphics*[width=2.7in]{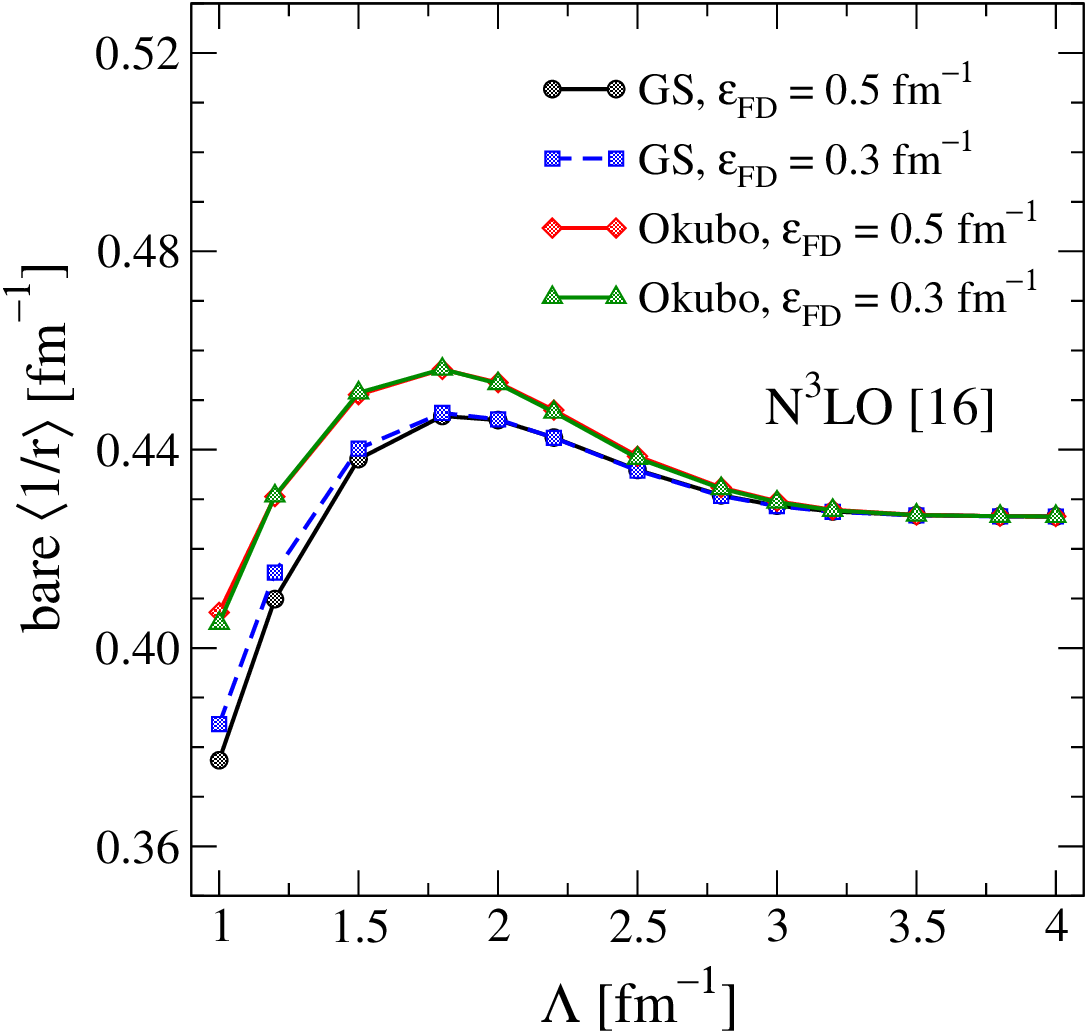}
\end{center}
\caption{The matrix element of $1/r$ in the deuteron
calculated with the bare operator as a function of the cutoff 
for different regulators and hermitization schemes.
The low-momentum interactions are derived using the N$^3$LO
chiral potential from Ref.~\cite{N3LO}.}
\label{fig:oneoverr}
\end{figure}

Matrix elements of operators that are dominated by distance scales 
larger than the inverse cutoff are to a good approximation preserved 
as the cutoff is lowered.
We investigate the evolution of operators by studying the expectation 
values in the deuteron of the bare quadrupole moment, rms radius and
$1/r$ operators. 
The relevant formulas are~\cite{Epelbaum:1999dj}
\begin{align}
Q_d &=  \frac{1}{20} 
\int_0^\infty \! dr \, r^2\, w(r)\, ( \sqrt{8} \, u(r) - w(r) ) \nonumber 
\\[2mm]
&= -\frac{1}{20} \int_0^\infty \! dk \, \biggl[
\sqrt{8} \, \biggl( k^2\, \frac{d\wt u(k)}{dk} \frac{d\wt w(k)}{dk}
+ 3 k\, \wt w(k) \frac{d\wt u(k)}{dk} \biggr) \nonumber \\[2mm]
& \hspace*{6cm}
+ k^2 \biggl( \frac{d\wt w(k)}{dk}\biggr)^2 + 6 \, \wt w(k)^2 \biggr] \,, 
\\[4mm]
r_d &= \frac{1}{2}
\biggl[ \int_0^\infty \! dr \, r^2 \, ( u(r)^2 + w(r)^2 ) \biggr]^{1/2}
\nonumber \\[2mm]
&= 
\frac{1}{2} \biggl[ \int_0^\infty \! dk \, \biggl\{ \biggl(
k \, \frac{d\wt u(k)}{dk} \biggr)^2 + \biggl(k \, 
\frac{d\wt w(k)}{dk}\biggr)^2 
+ 6 \, \wt w(k)^2 \biggr\} \biggr]^{1/2} \,,
\end{align}
and
\beqn
\langle 1/r \rangle  = \int_0^\infty \! dr \,
\frac{1}{r} \, [ u(r)^2 + w(r)^2 ] \,.
\eeqn
(We note that the momentum-space expressions for $Q_d$ and $r_d$ show
that these are
only well-defined for a smooth cutoff~\cite{Epelbaum:1999dj}.)
These operators are dominated by the long-range part 
of the interaction, and therefore we expect that the expectation
values change only for low cutoffs. This is verified in
Fig.~\ref{fig:Qd}, where we show the deuteron quadrupole moment $Q_d$ 
as a function of the cutoff for different smoothness regulators 
(left) and for different hermitization schemes (right), and in 
Figs.~\ref{fig:rsq} and \ref{fig:oneoverr}, 
where the rms radius $r_d$ and the matrix element of $1/r$ are plotted
as a function of $\Lambda$.

\begin{figure}[t]
\begin{center}
\includegraphics*[width=2.7in]{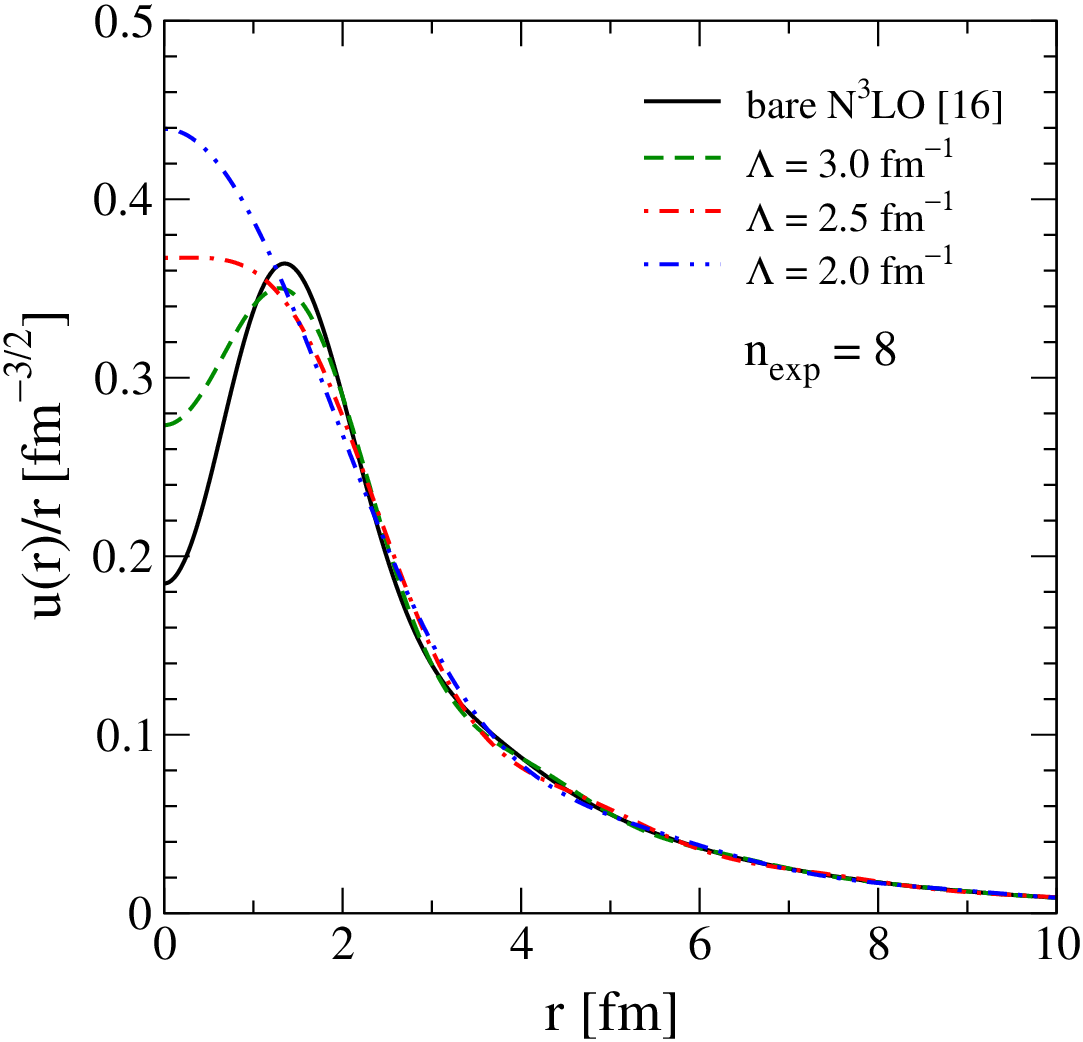}
\hfill
\includegraphics*[width=2.7in]{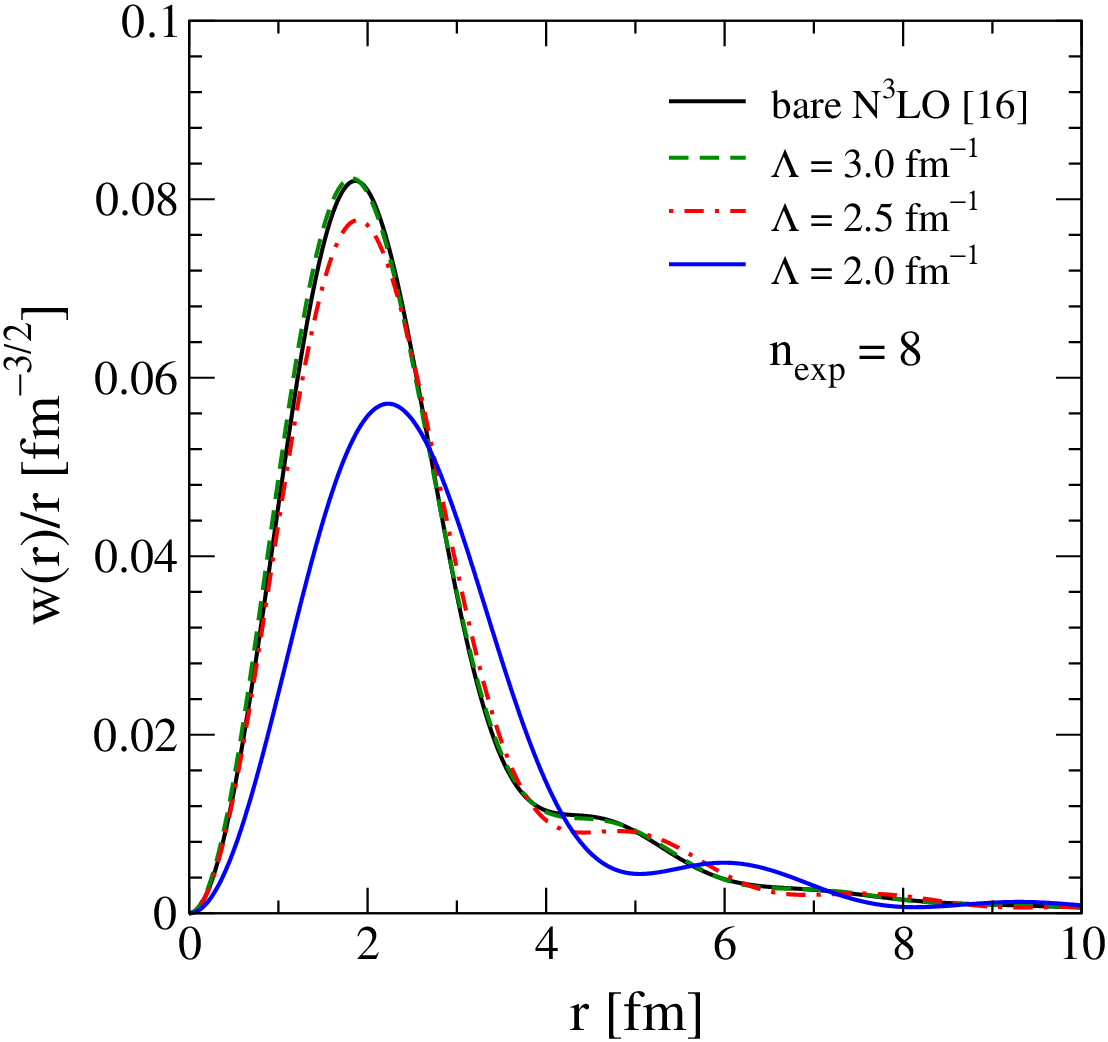}
\end{center}
\caption{Deuteron wave functions in coordinate space for low-momentum
interactions at several different cutoffs using an exponential regulator
with $n_{\rm exp} = 8$.  The initial interaction is 
the N$^3$LO chiral potential from Ref.~\cite{N3LO}.}
\label{fig:deuteronwfsp2}
\end{figure}

We emphasize that the bare quadrupole moment by itself does not
correspond to an experimental observable.
To correctly reproduce 
the experimental moment (indicated in Fig.~\ref{fig:Qd} by an arrow),
one needs the corresponding operators.
We find that the change of the quadrupole moment is of the same 
magnitude as the difference between the bare and experimental quadrupole 
moments for cutoffs down to $1.5$--$2\,\mbox{fm}^{-1}$. From the
deuteron wave functions shown in Fig.~\ref{fig:deuteronwfsp2},
we conclude that low-energy observables in the deuteron are 
reproduced without short-range correlations in the wave function.
Similar observations hold for the expectation values of the rms 
radius $r_d$ and of the $1/r$ operator.

\begin{figure}[t]
\begin{center}
\includegraphics*[width=3.2in]{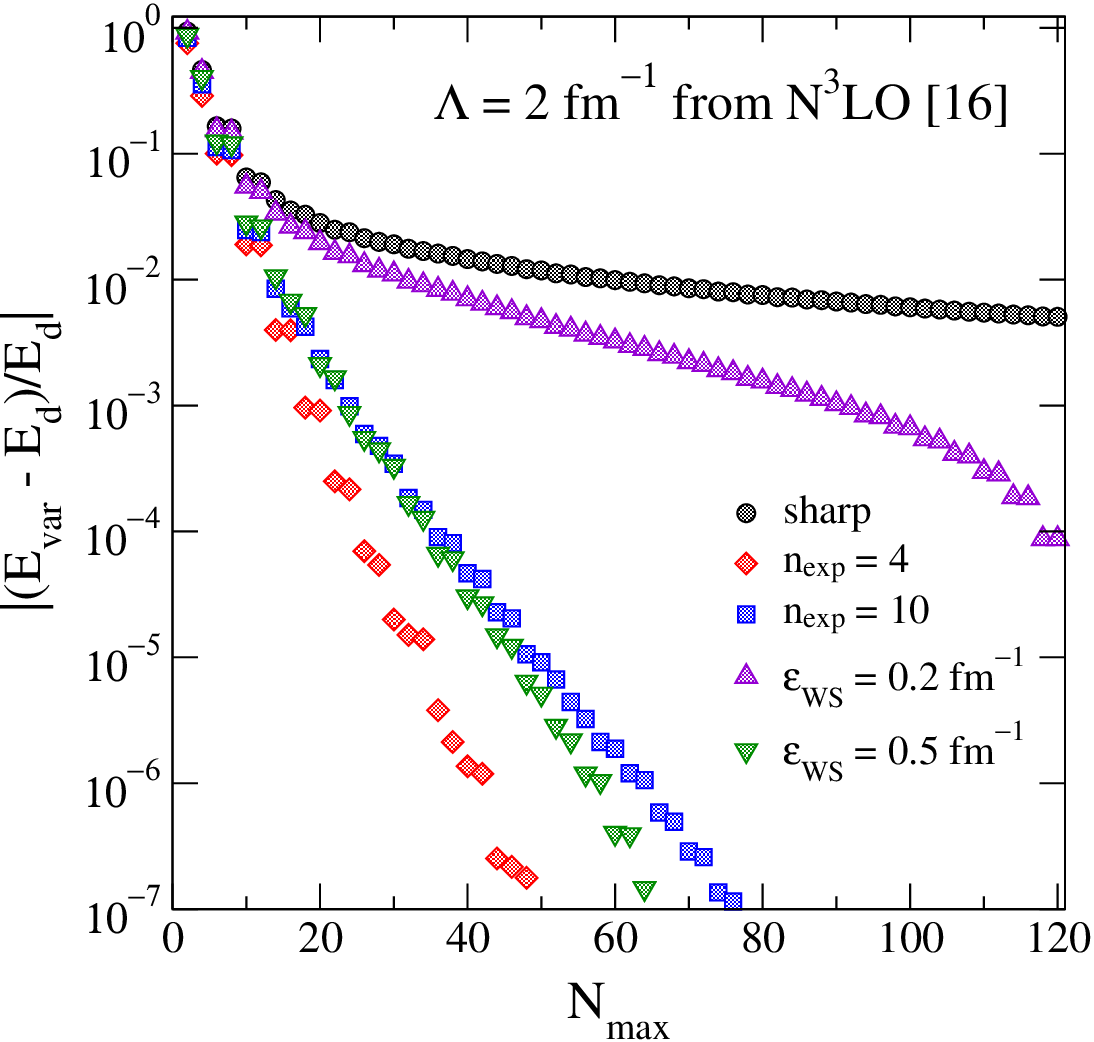}
\end{center}
\caption{The relative error in the deuteron binding energy $E_d$ as
a function of the size of the oscillator space for sharp cutoff
and various smooth regulators.}
\label{fig:deutconvergence}
\end{figure}

The key motivation for smooth cutoffs was to remedy the slow 
convergence at the 10\,keV level in the deuteron and at the
100\,keV level in the triton, when calculated in a harmonic 
oscillator basis.
In Fig.~\ref{fig:deutconvergence}, the relative error in the binding
energy of the deuteron (with respect to the converged 
result) is shown
as a function of the size of the oscillator space ($N_{\rm max} \, 
\hbar \omega$ excitations) for a range of regulators.
The slow convergence is evident for the sharp cutoff, where the error
is below the percent level only for the largest space.
The Fermi-Dirac regulator with $\epsilon = 0.2 \fmi$, which is still very
sharp (see Fig.~\ref{fig:regulators1}), gives improved errors but still
requires large spaces.  Increasing $\epsilon$ from $0.2 \fmi$
to $0.5 \fmi$ steadily
decreases the error until it improves rapidly with the space size, while
still only minimally distorting phase shifts.
Very similar relative errors are obtained for the exponential regulator
with $n=8$ (not shown) or $n=10$.  Lowering $n$ to $4$ reduces the error
still further, but at the cost of a potentially significant distortion
of the phase shifts.

\begin{figure}[t]
\begin{center}
\includegraphics*[width=2.7in]{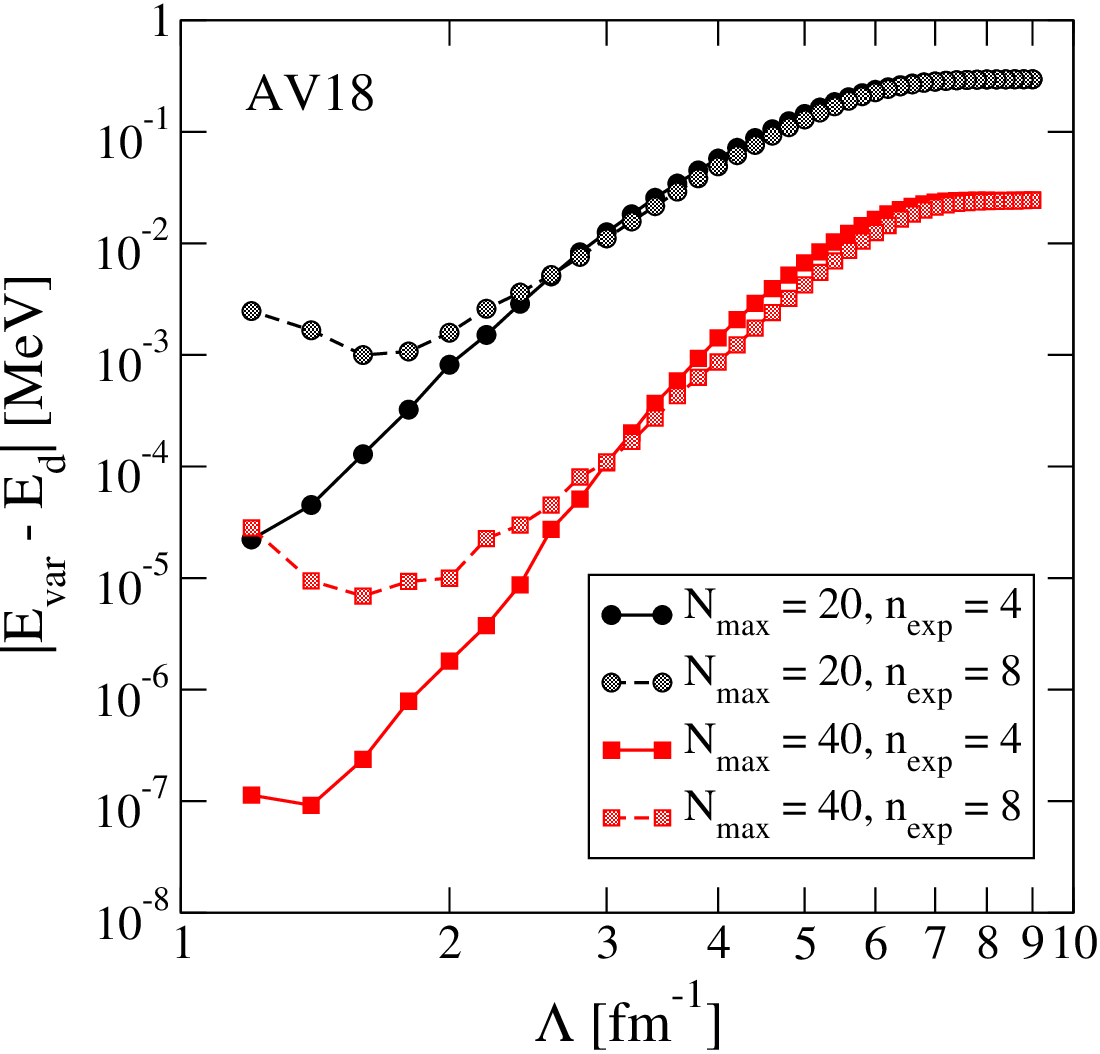}
\hfill
\includegraphics*[width=2.7in]{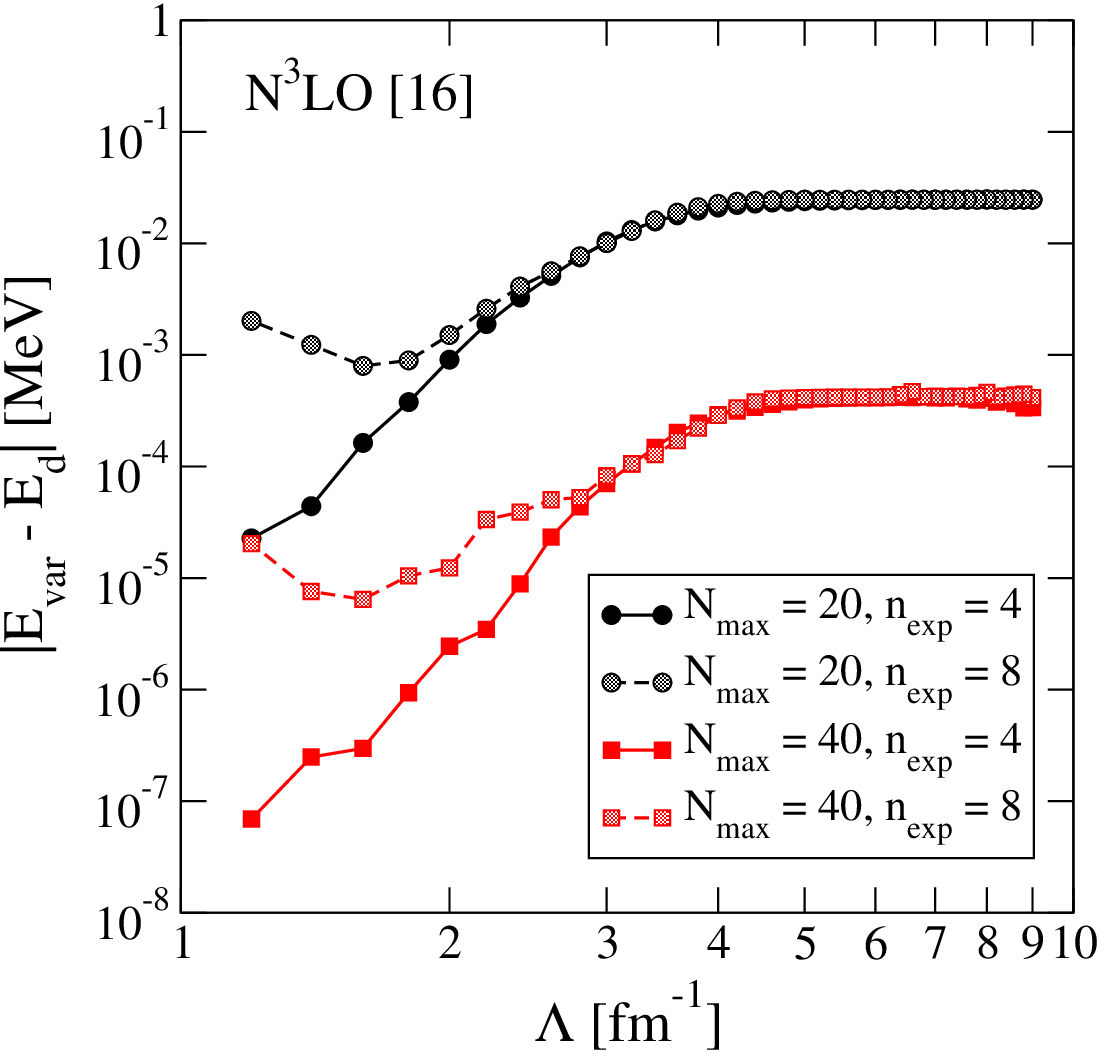}
\end{center}
\caption{The absolute error in the deuteron binding energy $E_d$ 
calculated from low-momentum interactions derived using the 
Argonne $v_{18}$ potential~\cite{AV18} (left) and the N$^3$LO 
chiral potential from Ref.~\cite{N3LO} (right) as a function of 
the cutoff. We show results for two oscillator spaces and for two
different exponential regulators.}
\label{fig:deutconvergence2}
\end{figure}

The absolute error in the deuteron binding energy is shown in 
Fig.~\ref{fig:deutconvergence2} as a function of the cutoff
for low-momentum interactions 
derived from the Argonne $v_{18}$~\cite{AV18} and the N$^3$LO 
chiral potential from Ref.~\cite{N3LO}. In all cases, 
the reproduction of the
binding energy improves with decreasing cutoff until the
limits of numerical accuracy are reached.
This is in contrast to the observed degradation below $2\,\mbox{fm}^{-1}$
for a sharp cutoff~\cite{Bogner:2005fn}. 
The improvement for the N$^3$LO potential starts
just below $4\,\mbox{fm}^{-1}$ (even though the EFT cutoff is
500\,MeV or $2.5\,\mbox{fm}^{-1}$) and is quite significant by 
$2\,\mbox{fm}^{-1}$.

\begin{figure}[t]
\begin{center}
\includegraphics*[width=3.2in]{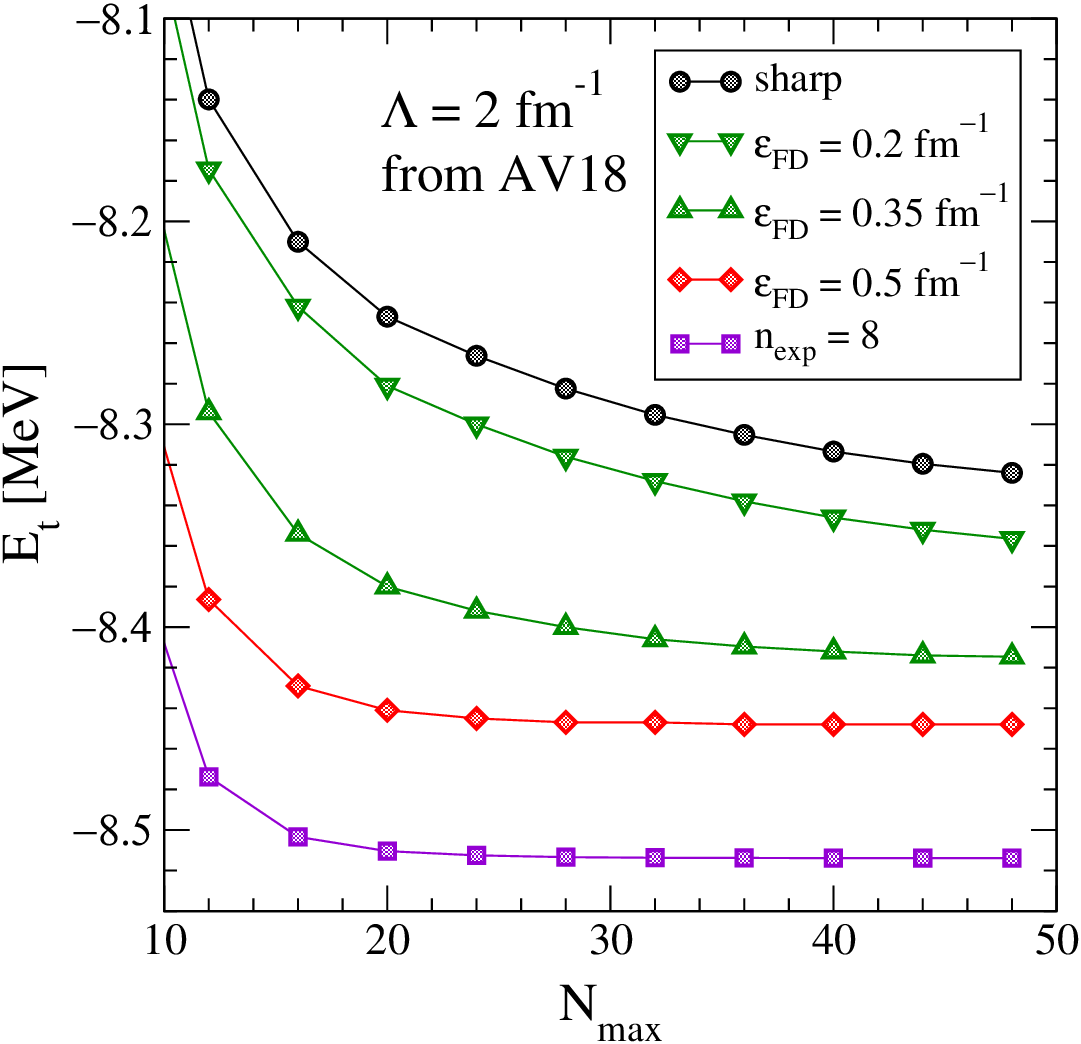}
\end{center}
\caption{The triton binding energy $E_t$ as
a function of the size of the oscillator space, 
for different smoothness regulators starting from the 
the Argonne $v_{18}$ potential~\cite{AV18}.}
\label{fig:triton2}
\end{figure}

The convergence is also greatly improved for the triton.
In Fig.~\ref{fig:triton2}, the triton binding energy is plotted as
a function of the size of the oscillator space.  For efficiency, convergence
for the smallest possible space is desirable, 
as the computational cost grows
rapidly with $N_{\rm max}$ and for larger systems.
Convergence at the keV level is achieved by all exponential
regulators with $n \geqslant 4$ ($n_{\rm exp} = 8$ is shown)
soon after $N_{\rm max} = 20$.
The consequence in moving from sharp to increasingly smooth regulators 
is seen from the Fermi-Dirac regulators, where $\epsilon = 0.5 \fmi$
yields very satisfactory results.
The difference between converged results for the triton
is a measure of the differences in the short-range three-body force
with the different regulators.

\begin{figure}[t]
\begin{center}
\includegraphics*[width=5.2in]{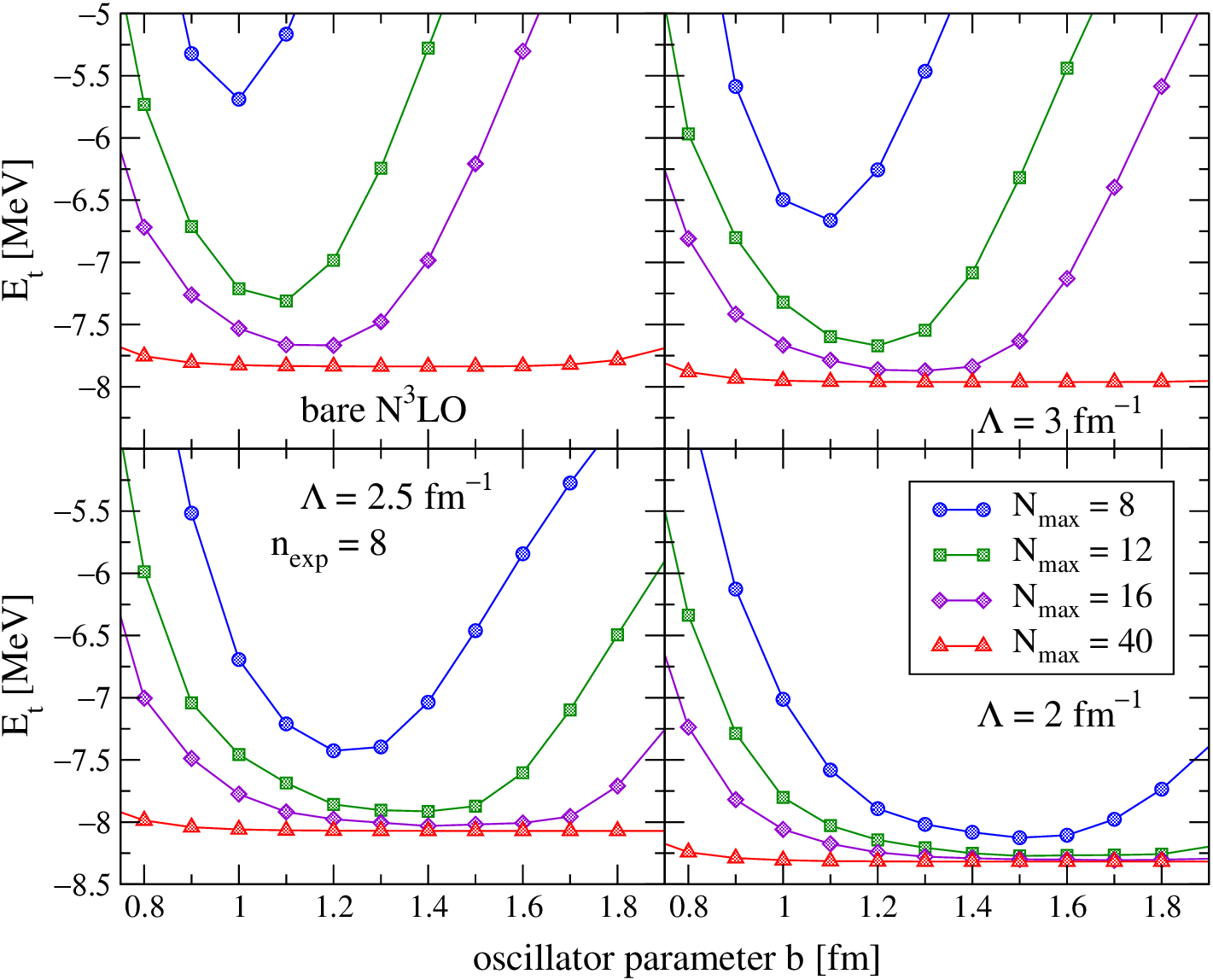}
\end{center}
\caption{Dependence of the triton binding energy $E_t$ as a function 
of oscillator parameter $b$ for the bare N$^3$LO chiral potential of
Ref.~\cite{N3LO} (upper left) and evolved to low momenta $\Lambda = 
3.0, 2.5$ and $2.0\,\mbox{fm}^{-1}$ with exponential 
regulator $n_{\rm exp}=8$. Results are presented for three smaller
spaces $N_{\rm max}=8, 12$ and $16$, along with the converged $N_{\rm 
max}=40$ energy.}
\label{fig:triton4}
\end{figure}

The dramatically
improved convergence of lower cutoffs compared to even the soft
N$^3$LO chiral potential is demonstrated in Fig.~\ref{fig:triton4}.
Note that the phase shifts would be essentially undistorted at 
$\Lambda = 2.5\,\mbox{fm}^{-1}$.
In addition, when we evolve the potential to low cutoffs, the
dependence on the oscillator parameter $b$ becomes flatter for 
a given $N_{\rm max}$, and for $\Lambda = 2.0\,\mbox{fm}^{-1}$ 
the minimum is unchanged as the space is enlarged. 
The sharp-cutoff $\vlowk$
also shows a flatter dependence on the oscillator parameter
compared to the initial N$^3$LO chiral potential and converges
rapidly to the 100\,keV level, but then converges 
slowly (the convergence for the N$^3$LO chiral potential is very similar
to that shown for the Argonne $v_{18}$ potential in Fig.~\ref{fig:triton2}).

\begin{figure}[t]
\begin{center}
\includegraphics*[width=3.2in]{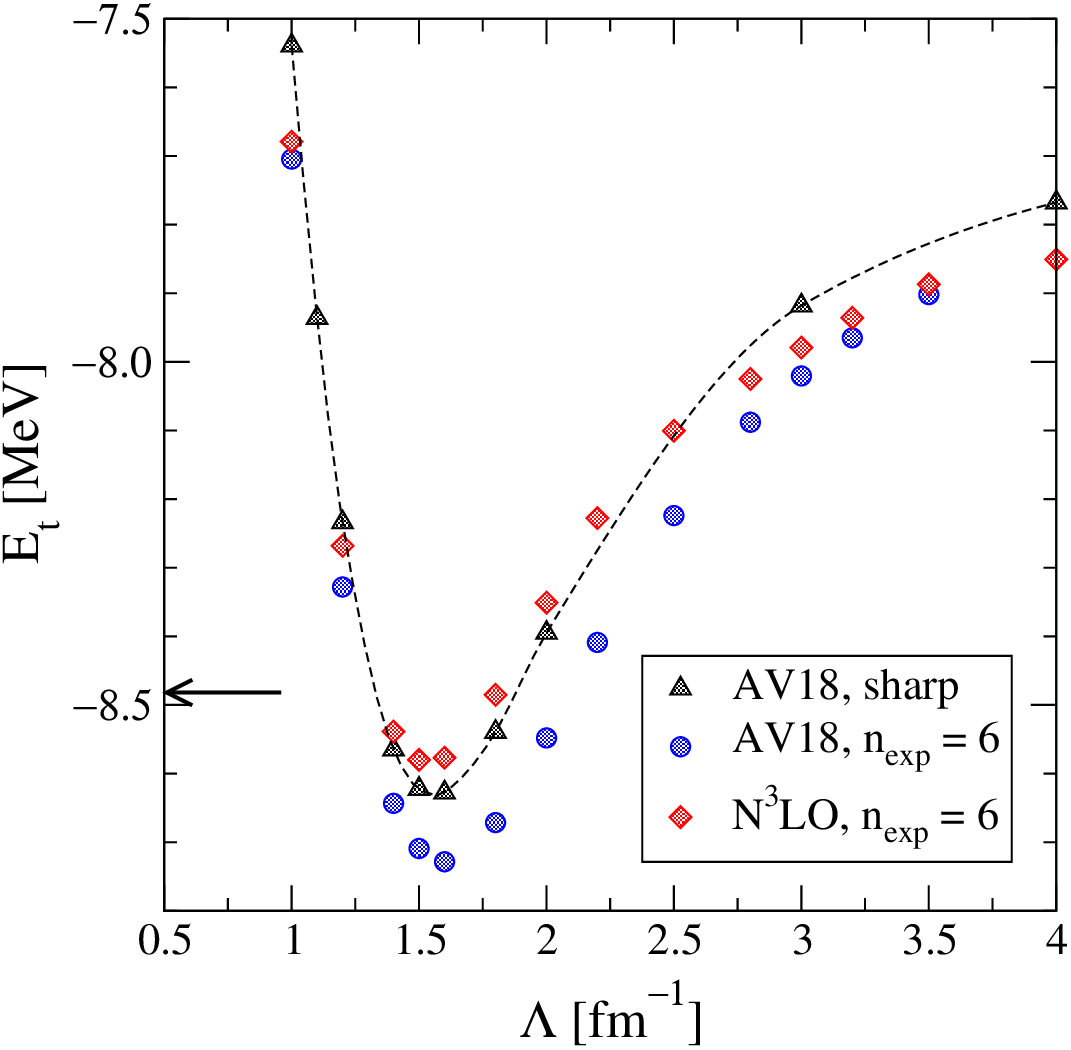}
\end{center}
\caption{The triton binding energy $E_t$ as a function of the cutoff
starting from the Argonne $v_{18}$ potential~\cite{AV18} with a 
sharp cutoff (taken from the Faddeev calculations of Ref.~\cite{Vlowk3N}) 
and with an exponential regulator. We also show results for the N$^3$LO
chiral potential of Ref.~\cite{N3LO} with the same exponential regulator.}
\label{fig:triton3}
\end{figure}

In Fig.~\ref{fig:triton3}, the triton binding energy as a function
of the cutoff is shown with the exponential regulator $n_{\rm exp}=6$
calculated from the Argonne $v_{18}$~\cite{AV18} and using the N$^3$LO
chiral potential of Ref.~\cite{N3LO}. For comparison, we also
show the Faddeev results from Ref.~\cite{Vlowk3N}, which use a 
sharp cutoff.
The triton binding energy is cutoff dependent because we have not included
the corresponding three-body interactions. Therefore, the difference to 
the experimental binding energy (shown with an
arrow) is the total three-body contribution.
The pattern of running is very similar in the three cases, although the 
values for a given regulator naturally vary.
While it is possible to choose a cutoff so that the experimental triton 
binding energy is reproduced by only the two-body interaction, the total 
three-body contribution will not vanish in other systems (e.g., the
alpha particle~\cite{Vlowk3N} or nuclear matter~\cite{Bogner_nucmatt}).


\section{Summary and Conclusions}
\label{sect:concl}

In this paper, low-momentum ``$\vlowk$'' interactions with smooth
cutoffs were
constructed and tested for several different types of regulators and
hermitization schemes. 
Problems seen in previous work with variational
calculations that were attributed to artifacts from a sharp cutoff are
resolved by smooth regulators.
In particular, convergence for the
energy of the deuteron and triton in a harmonic oscillator basis is greatly
improved. 
We have checked that this conclusion also holds for other variational
trial wave functions not shown here. 
We expect this improved convergence  to carry
over to other few- and many-body systems.

The regulators introduced here are specified not only by the 
cutoff on relative momenta but by a parameter that determines
the smoothness. Our conclusions for the optimal regulator are
provisional, because this work has been restricted to
convergence rates in the deuteron and the triton.
However, based on the observed convergence it appears that a similar
degree of smoothness is optimal for different regulators.
This is achieved with $n \approx 8$ for the exponential regulator or 
$\epsilon \approx 0.5 \fmi$ for the Fermi-Dirac regulator.
Going to smoother regulators maintains the same rapid convergence but
will increase distortions of the phase shifts. Whether or not this
is an issue depends on the application, but we expect that the
latter is not a problem for low-energy observables.

Given the regulator freedom that exists in generating low-momentum 
interactions for a given cutoff $\Lambda$, it is clearly 
a misnomer to speak of ``the'' $\vlowk$ potential.
We emphasize in particular that the consistent three-body 
(and higher many-body) interactions corresponding to the 
smooth two-body interaction will differ in each case.  
More precisely, since the long-distance (pion-exchange) parts 
of the interaction are preserved by the RG 
(until the cutoff is comparable to the pion mass),
the short-distance part of the three-body interaction is modified.
Cutoff and regulator independence with two-body interactions alone should
\emph{not} be a criterion for the applicability of low-momentum
interactions to nuclear structure.
(Note that this perspective differs markedly from that expressed
in Ref.~\cite{Fujii:2004dd}, which discussed the
application of low-momentum potentials with sharp cutoffs
to few-nucleon systems.)
Rather,
the three-body variations for different regulators allow us to
extend the powerful test of cutoff independence for observables 
to include independence of the regulator.
However, without a consistent
three-body interaction, few- and many-body calculations are
largely meaningless (except for considerations of convergence).  
Therefore, a high priority for the near term is to develop the 
machinery to quickly fit approximate three-body interactions 
for a given regulator.

The methods described here apply equally to low-momentum interactions
derived from conventional
nucleon-nucleon potentials and to those derived from chiral EFT potentials. 
The latter has the advantage of a systematic organization of
many-body forces and operators.
While chiral EFT potentials already start from lower cutoffs than conventional
NN interactions, we find significant added  
advantages for few- and many-body calculations 
by starting with chiral potentials fit at a larger cutoff
and running them down to a lower cutoff~\cite{Bogner_nucmatt,Bogner:2006tw}.  

Field redefinitions in EFT reshuffle higher-order terms 
so that the
truncation error is different but of the same order.  That is, if the EFT
is specified to NLO, then N$^2$LO terms will differ after a field redefinition
but the expected truncation error is still N$^2$LO.
The RG transformations discussed here preserve the error as the
cutoff is lowered by generating all necessary higher-order short-range
operators through the evolution.
This is particularly advantageous when the running is rapid
and the separation of scales is not large, as in the nuclear case (when the
tensor force is active). 
At present, only the two-body
interaction is evolved while the three-body interaction is fit for each
cutoff using the N$^2$LO form~\cite{Vlowk3N}.  
While there are solid indications that
this is a reasonable procedure~\cite{Vlowk3N}, the evolution of
consistent chiral three-body interactions to lower momenta is a major goal.  
In addition, the smooth regulators discussed here may have advantages for
constructing chiral EFT potentials with low cutoffs. As is apparent from 
Fig.~\ref{fig:regulators1}, the exponential regulator with $n \approx 8$ 
and the Fermi-Dirac regulator with $\epsilon \approx 0.5 \fmi$ lead to
weaker distortions than the conventional exponential regulator with 
$n = 3$ in the N$^3$LO chiral potentials~\cite{N3LO,N3LOEGM}, and thus
smaller errors have to be absorbed by the counterterms.

There are immediate applications that can be made using the low-momentum
interactions with smooth cutoffs developed here. Variational
calculations using hyperspherical harmonics~\cite{Viviani:2005gu}, 
the No-Core Shell Model~\cite{NCSM}
and the 
Coupled-Cluster approach~\cite{CC} should benefit from the improved 
convergence. Tests for all of these are in progress.

\begin{ack}
We thank J. Engel, E. Epelbaum, B. Friman, K. Hebeler, R. Machleidt, A. Nogga
and P. Vogel for useful discussions. 
This work was supported in part by the National Science 
Foundation under Grant No.~PHY--0354916, the US Department of Energy 
under Grant No.~DE--FG02--97ER41014, and the Natural Sciences and 
Engineering Research Council of Canada (NSERC). TRIUMF receives federal
funding via a contribution agreement through the National Research Council
of Canada.
\end{ack}


\end{document}